\documentstyle[prb,aps,epsf,epsfig,multicol]{revtex}
\newcommand{\be}{\begin{equation}}
\newcommand{\ee}{\end{equation}}
\newcommand{\ba}{\begin{array}}
\newcommand{\ea}{\end{array}}

\begin{document}

\draft

%2345678911234567892123456789312345678941234567895123456789612345678971234567898

\title{From Individual to Collective Pinning: Effect of Long-range Elastic
  Interactions}

\author{Anne Tanguy$^{+}$, Matthieu Gounelle$^{+,*}$} 

\address{$^+$ Laboratoire de Physique et de
M\'ecanique des Milieux H\'et\'erog\`enes, \\
URA CNRS n$^\circ$ 857,\\
Ecole Sup\'erieure de Physique et Chimie Industrielles de Paris, \\
10, rue Vauquelin, 75231 Paris Cedex 05,  France.}

\address{$^*$ Permanent Address:\\
Centre de Spectrom\'etrie de Nucl\'eaire et de Spectrom\'etrie de
Masse\\
B\^atiment 104, 91485 Orsay Campus, France.}
\author{St\'ephane Roux}

\address{Laboratoire Surface du Verre et Interfaces,\\
Unit\'e Mixte de Recherche CNRS/St-Gobain,\\
39 Quai Lucien Lefranc,\\
F-93303 Aubervilliers Cedex, France.}
\date{\today}

\maketitle
\begin{abstract} 
We study the effect of long-range elastic interactions in the
dynamical behavior of an elastic chain driven quasi-statically in a
quenched random pinning potential. This is a generic situation occuring in 
solid friction, crack propagation, wetting front motion, ... 
In the strong pinning limit, the dynamic of the
chain is controlled by individual instabilities of each site of the
chain. Long-range correlations in the
displacement field and in the force field develop progressively. The system
self-organizes to a steady state where the propagation of the instabilities is
described by scaling laws with characteristic critical exponents.
These exponents are numerically estimated through the analysis of the
spatio-temporal correlation in the activity map. 
Tuning the exponent $\alpha$ of the algebraic decay 
of the elastic interaction with the distance is shown to 
give rise to three regimes: a Mean-Field (MF) regime
valid for $\alpha<1$ (very slow decay), a Laplacian (L) regime for $\alpha>3$
(rapid decay of interactions), and an intermediate regime $1<\alpha<3$ 
where the critical exponents interpolate continuously between the MF and L
limit cases. 
The latter regime is shown to display, in the range $1<\alpha<2$, a
mean-field type character only for time-correlations but not for space. 
The effect of the driving mode on the avalanche statistics is also analyzed.
\end{abstract}

%%%%%%%%%%%%%%%%%%%%%%%%%%%%%%%%%%%%%%%%%%%%%%%%%%%%%%%%%%%%%
\begin{multicols}{2}
\section {Introduction}
Multistability of elastic media in a pinning potential is responsible
for the complex dissipative behavior observed in various physical
situations like motion of vortices in type II superconductors
\cite{Larkin}, dynamics of a ferromagnetic domain wall driven by an external 
magnetic field through a random medium \cite{Zapp}, Charge Density
Waves (CDW)\cite{Fisher}, roughening of crack fronts in
fracture \cite{Bouchaud}, or solid friction \cite{Caroli}. 

In the latter case, the dynamic of the slider can be reduced to that of 
the asperities at the
surfaces of the solid \cite{Mazur}. The role of the elastic body is to
mediate the interactions between asperities and with the pulling machine. 
It has been shown recently that the competition between the elastic restoring 
force due to the bulk and the non linear pinning force, due to the contact 
between asperities of different solids, gives rise to multiple stable 
equilibrium positions when the pinning forces are sufficiently strong, or 
when the system is sufficiently large. This multistability is responsible for 
hysteretic behaviour of asperities, when they are driven quasi-statically over 
the set of pinning centers belonging to the surface of the other solid. 
Dissipation arises from this hysteretic behaviour and it takes place in the solid, 
which plays as well the role of thermal bath. 
In the case of spreading of a partially wetting liquid on a heterogeneous 
plane\cite{Joanny},
the evolution of the contact line depends on the competition between a ``pinning 
force'' due to surface heterogeneities, and an elastic restoring force 
resulting from surface tension. In fact, a distortion of wavelength 
$\lambda$ of the three-phase front modifies the liquid-vapor interface over the 
same distance away from the wall, thus resulting, after integration over the whole liquid-vapor 
surface, in a effective elastic restoring force 
\be
\label{eq:inter}
f(x)=-k\int%_{\vert x-x_1\vert=a}^{L} 
\frac{u(x)-u(x_1)}{\vert
  x-x_1\vert ^\alpha}\, dx_1
\ee
with $\alpha=2$. The local capillary force is a non local function of the entire 
contact line position.
The same effect holds for crack propagation\cite{Gao}, where the stress intensity 
factor exhibits a non local dependence on the front geometry, with 
$\alpha =2$ as well. 
In the case of solid friction, the couplings due to the three dimensional elasticity 
are scale invariant and their translationaly invariant part has the same form 
(\ref{eq:inter}), with $\alpha=2D-1$ for a D-dimensional lattice of asperities 
coupled via a three-dimensional solid (thus $\alpha=1$ for $D=1$ and $\alpha=3$ for 
$D=2$, as can be checked by Fourier transforming the usual 
relations\cite{Landau}).
Expression (\ref{eq:inter}) is also compatible with dipolar interactions 
($\alpha =3$) in the case of ferromagnetic domain wall (D=2), and we will see that 
it also takes into account the usual Laplacian couplings of vortex lattices,
charge density waves ...

The effect of the long range coupling on the dynamics of the system has been 
little studied since each event involves a large part of the system and can
generally not be solved perturbatively. However, it is known\cite{Lee} to have a 
strong influence on the behavior
of the system as well as on its stability properties.
Here, we focus on the behaviour of an elastic line (D=1) with long range 
interactions, driven quasi-statically on a substrate with quenched disorder, 
and in the strong pinning limit.
We want to show the effect of the long-range elastic couplings on the 
fluctuations accompanying the average behaviour of the line.

By analogy with friction, we will call ``asperity'' each point of discretization 
of the elastic line. In general, two situations may hold: if the 
couplings are small in
comparison to the distortions of the pinning potential, then
the motion of the chain is controlled by the motion of each
asperity. This is the {\em strong pinning} limit. In this case, when the local
stiffness is sufficiently weak, the stable local equilibrium position
of an asperity may disappear, and the ``unstable'' asperity advances
suddenly to the next equilibrium position. In contrast, in the {\em weak
pinning} limit, when the
couplings are sufficiently strong, the line behaves in a first
approximation as a whole. In such a case,
a separation in domains may occur for a sufficiently large
system size\cite{Larkin}.  Thus a coarse-graining of the system
at large enough scales leads to a simple strong-pinning regime.
The latter case thus appears to be the possible relevant limit in the 
thermodynamic limit of a infinite sized system.

In this paper, we investigate the effect of the range of elastic interactions 
in the {\em strong pinning case}, where the fluctuating dynamics of the system 
results from the deterministic propagation of local instabilities. 
More precisely, we focus our attention on the interactions between asperities 
and we consider only the elastic displacement field components whose 
wavelength is greater than the distance between ``asperities''. This approach is 
complementary to that of considering individual jumps without couplings\cite{Joanny}.
We show how the nature of the elastic interactions allows to interpolate 
between a ``delocalized'' (or mean field) regime and a
``local'' regime analogous to the Laplacian case. 
The driving mode is shown to affect the velocity of propagation of
the instabilities and the avalanche statistics.

In the first part of the paper, we propose a discrete model of
the dynamics which will be compared to the continuous one. Then we study in
a systematic way the dependence on the decay of elastic interactions,
$\alpha$ in Eq.(\ref{eq:inter}), of the
propagation of instabilities, of the organization of the dynamic and
of the kinetic roughening. The last section is devoted to a discussion
about the universality of our model, and on the uniformization due to
elastic interactions.

Let us recall briefly that $\alpha=1$ for an elastic
one-dimensional chain with ``three-dimensional'' couplings, but
$\alpha=-2$ for an elastic string with bending elasticity\cite{Landau},
$\alpha=2$ for fracture dynamics in infinite solids\cite{Gao} or wetting 
experiments in a free surface geometry\cite{Joanny}, and $\alpha=0$ in propagating 
capillary fronts due to the fluid flow behind the front\cite{Webman}.

It is important to note that, for a given physical system, $\alpha$ may be 
changed experimentally by changing the geometry of the apparatus. This is, for 
example, the case of wetting, where $\alpha=2$ in a free surface geometry, but 
where the couplings are Laplacian at large scales in a Hele-Shaw cell (at 
vanishing capillary number)\cite{Paterson}.

\section {The model}

Our aim is to single out the effect of interactions in the propagation of
instabilities in an elastic chain (D=1) driven quasi-statically along its 
transversal direction over a rough surface. 
We consider here a discrete one dimensional elastic chain of size
$L$. The chain of size $L$ is discretized into $L/d$ blocks. $d$ represents the
distance between asperities in the strong pinning limit, and more generally
the distance between ``sites''. In the case of wetting for example, it will be 
at least on the order 
of the distance between pinning centers. It is not possible to have access to 
information below $d$ in our model; $d$ is the lower cut-off of our model and 
$L$ is the upper cut-off.

We call $\bf x$ the current site which is the position of the site 
evaluated in the non deformed geometry. The position of the site in the 
deformed geometry is $\bf{x'}=\bf{x}+\bf{u(x)}$, where $\bf{u(x)}$ is the 
displacement field of the chain in the laboratory frame. We will refer as well 
to $\bf{u(x)}$ as the difference in the actual position $\bf{x'}$ of site $\bf{x}$ to any 
homogeneous motion of the undeformed geometry.  
We consider only the large wavelength components of the displacement 
field ($\lambda >d$), thus $\bf{u(x)}$ corresponds to the displacement averaged over 
a cell of size $d^D$. In the following, we take account only for the 
displacement in the direction of pulling, $\bf j$, transversal to the chain, 
and we neglect the displacement along the chain. The displacement field is a 
scalar and it will be noted $u(x)$, as shown in Fig. (1). 

Each site of the chain is subjected to 
a driving force, a random pinning force (from the interface) and a long range 
elastic force describing the coupling to the other sites of the chain.  
The continuous equation of motion for this system could thus be written in the 
overdamped limit

\begin{eqnarray}
\gamma \frac{\partial u(x,t)}{\partial t} &=&
\bigg[F_{ext}(t) + \eta
(x,u(x))\label{eq:mvt}\\
&& -k \int \frac{u(x)-u(x_1)}{\vert
  x-x_1\vert ^\alpha} \, dx_1\bigg] \nonumber 
\end{eqnarray} 
where $F_{ext}$ stands for the driving force and $\eta(x,u(x))$ for the 
quenched pinning force.
The left-hand side of the equation contains a phenomenological damping term.

The long range elastic force has been already introduced in the previous part 
of this article. Let us remark that it may be described by three fundamental 
properties: (1) it is linear in $u(x)$ (linear elasticity), (2) it is scale 
invariant (despite the lower cut-off $d$ and the upper cut-off $L$), (3) and it 
is ``translationaly invariant''. This latter property is due to the fact that we
call ``coupling forces'' only internal forces. These forces do not contribute 
to the average over all the medium. This is compatible with the fact that the 
elastic 
energy of the chain (in absence of pinning) is translationaly invariant. Any 
uniform displacement of the chain does contribute to an {\em average} restoring 
force which has to be compensated at equilibrium by a pinning force; but 
contrary to the definition used by C.Caroli et al.\cite{Caroli}, it does 
not contribute to the elastic {\em couplings}.
These couplings have an infinite range due to their algebraic decay. 
For the sake of simplicity, we will refer to $\alpha$
as the {\em range} of the interaction. As $\alpha$ increases, the
interaction tends to be concentrated dominantly on the nearest neighbors.
On the contrary for a small $\alpha$, the interaction tends to be
more evenly distributed over the system.
We will see later how to adjust the external driving force to take account 
for the average dynamic of the system. 
The quenched pinning force, $\eta$, represents the interaction between the 
heterogeneous surface and the elastic chain. In case of solid friction, it 
may have various physical origins (Hertzian contact between asperities of 
different surfaces, adhesion ...) but it is 
always non linear. Moreover, a large scale description of the contact problem 
supposes that the basic interaction is already the combination of multiple 
elementary processes, giving rise to discontinuities and multivaluation\cite{Caroli}. 
At a sufficiently coarse-grained level, the effective force of the interaction will 
loose its continuity.
In order to be able to describe the possible fixed point of a renormalized 
description of the surface interactions, we choose to use a discontinuous 
pinning force, described only by its statistical properties. This force is 
quenched and random. We will restrict ourself in the following to white noise.
Moreover, we anticipate that the fluctuations of the amplitude of the pinning 
force are 
small when the slider moves forward, and not sufficiently high to allow 
backward motion of the slider. This is also the case when {\em spatial} distortions 
of the pinning forces are strong. In this case, it is still
possible to use the expression (\ref{eq:mvt}) but with the reservation
that only the {\it positive} part of the r.h.s. is considered. 
Thus, the system is supposed to move only
forward and the pinning force may be described for each site $x$ only by a position 
coordinate $i(x)$ and a threshold value $\eta _{i(x)}(x)$ as shown in Fig. (1.b).
In such a manner, our model contains the main features of the dynamics of an 
elastic chain, driven over a pinning surface in the {\it strong pinning} limit.
It is also believed to capture the large scale behavior of any pinning 
potential. But, considering deliberately
the strong pinning regime, we cannot address the interesting question of the
cross-over scale from weak to strong pinning.

We are interested in the quasi-static limit of the motion of the chain. 
In this case, the damping is supposed to be sufficiently strong to allow each 
site to reach instantaneously its local equilibrium position. In the 
quasi-static approximation, it is well known that, in the stationary regime and 
in the strong pinning limit, the main contribution to the global displacement
is due to hysteretic jumps of each element of the chain, resulting from the
multistability of the site and of the delay rule\cite{Tang}. For a given 
external driving, the initiation of a jump is a local instability and thus the 
dynamics cannot be controlled simply by that of the center of mass. 
To avoid the explicit introduction of the dynamic of the system (with inertia 
and viscous damping), and thus the necessity to integrate the equation of 
motion, we assume the following conditions.
First, the system evolves from one local equilibrium position to the next 
closest one.
Second, we monitor the driving force $F_{ext}(t)$ actively so that we always 
maintain the system at the edge of stability. More precisely, $F_{ext}(t)$ 
is adjusted so that only one mode is metastable (only one site jumps at each 
time step).
Let us emphasize that in contrast to many approaches based on continuum 
equations as Eq.~(\ref{eq:mvt}), we do not consider a constant force but 
rather a constant but infinitesimal mean velocity.
In solid friction, it corresponds to a uniform average imposed displacement 
$\bf v$ on the top side of the solid. In wetting experiments, it will 
correspond to motion at an imposed slow rate of flow. 
Alternatively, $\bf v$ may be deduced from a motion where a force is imposed 
on the top side of the solid (given pressure in wetting experiments). $\bf v$ 
is related, via shear of the solid, to the average position of asperities of 
the chain.
When the displacement is imposed, the discrete version of our model, as 
studied below, belongs to the class of extremal models (such as invasion 
percolation\cite{WW}, fracture models \cite{HR(chap.5-6)} or more
recently introduced growth models\cite{BS}). With the second assumption,
the motion will consist in a series of equilibrium points, and the equation
of motion does not have to be integrated. Under the assumption of
the over-damped nature of the motion, we will discuss a possible extrapolation
of the obtained results to different driving modes.  This discussion
will allow us to relate the avalanche statistics to the observable
intermittency of the motion of the chain.

The above introduced model is discretized to allow for numerical simulations.
The string is described as a one-dimensional, periodic array of sites. The unit 
length along the chain is given by the distance between the sites of the chain, 
and the unit time by the time step $\delta t$ ellapsed between subsequent jumps. 
Our ``time'' count indeed the number of jumps (that is the main travelled 
distance). When one site depins from one 
asperity, it jumps by a distance $\delta u$, whose maximum value gives the unit length
in the direction of propagation. The pinning potential is described by a set 
of discrete centers, with a random spatial distribution. We assume that 
the amplitude of each pinning force is identical throughout the medium.  
Alternatively,
we could have chosen a uniform distribution (on a regular lattice) but
with a random magnitude.  We checked that these two variants (and their
combination) give the same results, as soon as some randomness is introduced.
For a periodic array of pinning centers with the same amplitude, as
could have been guessed, the chain motion finally locks on a periodic
motion of little interest.

The model runs as follows.
We start with a uniform flat front, $u(x)=0$, for each cell. We suppose 
that these positions correspond to equilibrium positions in absence of external 
driving force (that is for each $x$, $\eta(x,0)=0$). The external driving force
$F_{ext}$ is applied progressively. The elastic displacement is negligible, 
thus only the pinning force increases first ($\eta(x,u(x,t))=-F_{ext}(t)$).
When the load is sufficiently high, one cell overcomes its threshold, and 
jumps. The maximum load is in this case
$$F_{ext}= {\rm min}_x(-\eta_0(x)).$$
The cell $x^*$ which overcomes its threshold then jumps to the next basin 
$i(x^*)$, thus $$u(x^*)\rightarrow u(x^*)+\delta u$$
During the jump, the external driving force is constant. The linearity of the 
problem allows us to calculate the new elastic couplings. The elastic forces 
along the chain are modified to
$$f(x)\rightarrow f(x)+\delta u.G(x,x^*)$$
where $G(x,y)$ results from the interaction kernel discussed above.
The external driving force is then reduced to zero in order to prevent another 
jump, and the external load may be increased until 
$$F_{ext}= {\rm min}_x(-\eta_{i(x)}(x)-f(x))$$
in order to allow the next jump.
The discrete interaction kernel is chosen as the simplest expression which 
captures the required periodic boundary condition, and follows the 
power-law decay of Eq.~(\ref{eq:inter}):

\begin{eqnarray}
G(x,x^*)&=&\left[\sin\left(\pi\frac{\vert x-x^*\vert}{L}\right)\right]^{-\alpha}
\quad {\rm for}\ x\neq x^*
 \label{eq:gij}\\
 G(x^*,x^*)&=&-\Sigma_{x\neq x^*} G(x,x^*)
 \label{eq:sij} 
\end{eqnarray}
In the particular case of $\alpha=2$,
such an expression (\ref{eq:gij}) is the exact result of the summation of 
interactions for a periodic chain. 

This describes one elementary step of the model.  We start from a
uniform flat front and run a long sequence of such steps until the
system reaches a statistically stationary state.  
Condition Eq.~(\ref{eq:sij}) stands for the average conservation of forces 
along the chain. We used $G(x^*,x^*)\equiv 1$, thus the stiffness for 
neighboring sites is unity.
This sets the scale for forces. The jump size $\delta u$ thus reflects simply the
distance between asperities on the track \cite{Tang}. 
The dynamic is ``extremal'' in the sense that only the site submitted
to the maximum force $f(x^*)$ advances at each time step. 

Fig.\ref{fi:acti} shows examples of the space time distribution of ``active
sites'' $x^*$ in the medium for various couplings, $\alpha$.
Fig.\ref{fi:depla} shows the time evolution
of the displacement front along the chain for the same values of $\alpha$. It
appears clearly from these figures that the system organizes in a statistically 
stationary state, and that for large $\alpha$ (local couplings), 
the evolution is spatially inhomogeneous.

We compared several statistical distributions for $\delta u$ e.g. uniform in
the interval $[a,1]$. Our results at sufficiently large times and scales
did not display any dependence on $a$ as long as $a\neq 1$.  As noted above
when $a=1$, the pinning centers are periodically distributed, with the
same strength, and the chain motion ends up being periodic with no
interesting features.

Similar models have already been studied \cite{Schmitt} in the case $\alpha=2$ 
and in the limit as $\alpha \rightarrow\infty$ which is the discrete
version of the Laplacian one \cite{Hansen,Maslov} and we will refer to
them in the following.  Similarly, the case $\alpha=0$ corresponds 
clearly to a mean-field situation and it admits a simple analytical solution 
which is quite similar to the situation solved by Flyvbjerg et
al\cite{Flyvbjerg}.

As far as one is interested only in the effect of the long range
elastic interactions, it appears convenient to consider $\alpha$ as a
continuous parameter of the model, albeit in most physical cases, only integer
values can be found. 

\section {Kinetic roughening}

Starting with an uncorrelated distribution of forces, the system
organizes, after a transient, into a highly correlated statistically
stationary state. One manifestation of these long range correlations
is the roughening of the displacement front of the chain. The
roughness of the chain is characterized by the scaling of the
correlation function 
\be\label{eq:sawin}
\langle (u(x+\vartheta)-u(\vartheta))^2\rangle_\vartheta\propto x^{2\zeta}
\ee
or in Fourier space
\be\label{eq:safou}
\langle\vert\tilde{u}(k)\vert^2\rangle\propto k^{-1-2\zeta}
\ee
Such a power-law behavior of the power spectrum of the front position
is shown on Fig.~\ref{fi:rugo} for $\alpha=2$. 
$\zeta$ is called the ``roughness exponent'' of the front. 
From Fig.~(\ref{fi:rugo}) we determine $\zeta=0.35\pm 0.02$ in good agreement
with a previous determination of this exponent by Schmittbuhl et al \cite{Schmitt}.  

As can be
seen in Fig.~(\ref{fi:zetalpha}), $\zeta$ depends on $\alpha$. The larger $\alpha$,
the larger the roughness exponent, i.e. the more persistent the front fluctuation.
Note that for large
$\alpha$, the roughness exponent exceeds one, and thus Eq.~(\ref{eq:sawin})
is inappropriate.  Either one should use such a correlation function computed on 
the {\em slope} of the front (and measures $\zeta-1$), or revert to the 
spectral method for determining the roughness exponent.
For $\alpha=3$, we obtain $\zeta\approx1.2$ again consistent with previous
determinations \cite{Hansen}. The latter case, i.e.\ for local couplings, 
is equivalent to the Edwards-Wilkinson growth model\cite{}, however, the 
quenched nature of the noise induces a major change in the roughness exponent 
$\zeta$, from $\zeta=0.5$ for an annealed noise to $1.2$ for a quenched noise.

Fisher et al. \cite{FisherII} proposed a renormalization group analysis
of this problem and obtained 
\begin{equation}
\zeta_F=\frac{2\alpha-3D}{3}\label{eq:fish}
\end{equation}
where $D$ is the space dimension, i.e. in our case $D=1$. 
Expression (\ref{eq:fish}) has been obtained by replacing the
Laplacian propagator $1/q^2$ by $1/q^{\alpha-D}$ in the original
calculation in order to take into account the long-range coupling. 
This theoretical prediction is plotted as a dotted line in 
Fig.~\ref{fi:zetalpha}.  We observe a fairly good agreement for 
$1<\alpha<3$. However, we observe a marked difference for the 
Laplacian case, where the theoretical prediction $\zeta_F=1$ lies 
outside of the error-bars of our measurement.

The ``roughness exponent'' can be measured experimentally.  
In the case of wetting on heterogeneous surfaces in a Hele-Shaw
geometry,  A. Paterson et al. \cite{Paterson} reported
$\zeta=0.77$. Above a scale equal to the aperture of the cell,
the problem should be described by the case $\alpha=3$.  However, 
gravity plays a significant role in this problem introducing an additional 
confining term cutting down large scale front fluctuations.  
It is interesting to note that when gravity effects are suppressed (horizontal cell)
the geometry of the invading fluid is similar to that of invasion percolation.
The relation between invasion percolation and the quenched Edwards-Wilkinson 
problem has been discussed by Roux and Hansen\cite{Hansen}, in relation with
a model of this process proposed by \v Cieplack \cite{Cieplack}.
It may be that the expected scaling is restricted to a rather narrow window
limited from below by the cell thickness, and from above by the
capillary length
(measuring the effect of gravity as compared to the surface tension).
In case of wetting in a free surface geometry ($\alpha=2$), E. Rolley et al. 
\cite{Rolley} have found $\zeta\approx 1/3$, in very good agreement with our 
numerical result ($\zeta=0.35$). Contrary to the explanation proposed by Robbins
 et al.
\cite{Joanny2}, using Imry and Ma arguments, our explanation takes into account 
the local dynamics of the chain, and may be extended by the same way to any 
range of the couplings.
Another experimental situation which might be compared to our computation 
is the propagation of an interfacial crack.  Such a situation between two 
PMMA plates has been studied recently by Schmittbuhl et al \cite{SchmittII}.
They measured $\zeta\approx 0.55$, a value which is much higher than the 
above mentioned value $\zeta=0.35$ for $\alpha=2$.  The visco-elastic 
mechanical behavior of PMMA, or the initiation of cracks ahead of the front
which merge with the front giving rise to a tortuous geometry where
higher order terms may be relevant, may be responsible for this difference.
Not also that Ref.\cite{bouchaudII} is irrelevant in our case because
the crack is not interfacial there.

\section {Dynamical Organization}

Starting with an uncorrelated distribution of forces, the system
organizes after a transient into a correlated statistically
stationary state. In particular, the memory of the initial state is lost and
the distribution of forces $f(x)$ reaches a stationary distribution shown
in Fig.~(\ref{fi:fc}). 
The forces have a critical value $F_{max}$ depending
on the value of $\alpha$. The distribution is asymetric. 
The higher the value of $\alpha$, the more peaked is the distribution 
close to $F_{max}$. 
This distribution depends on the spatial correlations present along the front
and on the distribution of force increment following an elementary move.
The latter aspect can be taken into account \cite{tangroux} in a mean-field 
model, expected to hold for $\alpha<1$ as shown below.
The maximum depinning force $f(x^*)$ as well (related to the external force by 
$f(x^*)=-\eta_{i(x^*)}(x^*)-F_{ext}(t)$) has a lower critical value
corresponding to the   
maximal {\em external} force for the whole system, $F_c$. 
For a uniform constant 
depinning threshold for example( $\eta_{i(x^*)}(x^*)\equiv \eta_c=-1$), the lower critical 
value of the maximum depinning force is ($1-F_c/L$). 
This has been drawn in Fig.~(\ref{fi:fc}). $F_c$ increases with  $\alpha$.

\section{Propagation of instabilities}

\subsection{Spatio-temporal map of activity}

As shown in Fig.~(\ref{fi:acti}), the active sites are spatially and
temporally correlated.  These correlations can be trivial as when
$\alpha=0$, or more complex for $\alpha>1$.
In order to analyze these correlations for
various values of $\alpha$, it is of interest to study the probability
distribution\cite{Furuberg} $p(r,\Delta t)$ of having a distance $r$
between the sites
active at time $t$ and $t+\Delta t$ (i.e. the spatial
distribution of the $\Delta t^{{\rm th}}$ active site). From the numerical
simulations, we
observe that it is possible to describe the entire dependence of
$p(r,\Delta t)$ for $\alpha >1$ through a scaling form
\be
\label{eq:fur}
p(r,\Delta t)=\Delta t^{-1/z}\phi\left(\frac{r}{\Delta t^{1/z}}\right)
\ee
with a dynamic exponent $z$ which describes the spreading of the activity
with time over a zone of size $\xi\propto\Delta t^{1/z}$.
For $\alpha\rightarrow\infty$, $z=2$, as for a diffusive system.
Fig.~(\ref{fi:Furuberg}) shows a data collapse for $\Delta t$ ranging
from 1 to 64 and for
$\alpha=2$. The exponent $z$ is determined from this collapse and it
varies continuously with $\alpha$. The scaling function $\phi$
displays the following behavior:
\begin{equation}
 \phi(x)\propto\cases{ x^{-b} &for\quad $x\gg 1$\cr 
                       x^0 &for\quad $x\ll 1$\cr}
\end{equation}
For distances larger than the active length $\xi$, the decay of the
scaling function is characterized by an exponent $b$ which also depends
on $\alpha$.
Fig.~(\ref{fi:bzalpha}) summarizes the dependence of $b$ and $z$ on $\alpha$.

Let us first propose some arguments that allow us to understand
the different observed regimes as a function of $\alpha$.

First, it is possible to link the dynamical roughness exponent $\zeta$ to the
dynamical exponent $z$ in our model. Let us consider a starting point
$(x_o,t_o)$.  After
a time $\Delta t$, the activity as spread over a distance $\xi(\Delta
t)$ around $x_o$. The number of moves which have been necessary to
cover the area between the crack fronts at time $t_o$ and
$t_o+\Delta t$ scales as $\Delta x.\Delta u$. The front has a 
self-affine geometry with a roughness exponent $\zeta$, hence
$\Delta u\propto \Delta x^{\zeta}$.  As a result, we get the scaling of
the time difference with the extent $\Delta x=\xi$ as 
\be
\Delta t=\xi^{1+\zeta}
\ee
hence
\be\label{eq:zz}
z=1+\zeta
\ee
This scaling relation is accurately obeyed in our numerical simulations 
as long as $0\le\zeta\le 1$.  For larger values of $\zeta$, the above scaling 
breaks down, and $\Delta u\propto \Delta x$, hence, 
the effective exponent $z=2$ appears. 

Second, if we make the assumption that in the steady state there is only weak
spatial distortions in the force distribution, then $p(r,\Delta t=1)$ should
essentially reflect the load sharing rule due to the elastic
coupling. This argument simply predicts 
\be\label{eq:ba}
b=\alpha
\ee
as found in the numerical simulations for $\alpha<3$.
An analogous relation has been found in the propagation of initially
localized perturbation in the elastic map, without pinning
\cite{Torcini}. 

Third, as $\alpha$ increases, the load redistribution is much higher for the 
nearest neighbors than for the rest of the chain.  More precisely, when $\alpha=3$
the interaction force 
$$
f(x)=-k\int {(u(x)-u(x'))\over|x-x'|^3} dx'
$$
contains a singular part proportional to 
\be
f_{sing}(x)\propto \left({d^2u\over dx^2}\right) 
\ee
plus a regular part.  The same holds for any higher value of $\alpha$, up to 
$\alpha$ tending to infinity where $f(x)\propto d^2u/dx^2$.  Hence, the dynamics 
of the chain is essentially controlled by this singular part, and thus is no longer
expected to depend on $\alpha$ but rather should be equivalent to the case 
$\alpha=3$ or $\alpha=\infty$, i.e. the simple Laplacian (local) kernel, or the
Edwards-Wilkinson equation with quenched noise universality class.  This regime
will be referred to as Laplacian, or ``L'' regime, in the following.

Let us now return to distribution of distances between active sites 
at a time interval $\Delta t$. We observed the existence of a region of extent 
$\xi\propto \Delta t^{1/z}$ centered on the initiation site where most of the 
activity is confined, we call this region ``cluster'' ---although it is 
not connected in the space-time map. These clusters have a self-affine structure.  
Their scaling is indeed given by $\Delta x\propto\xi\propto \Delta t^{1/z}$.
We note that when $r>\xi$, the last occupied site is not part of the cluster
which originates on the former site.
The way the activity is distributed in space and time is controlled by
the statistical distribution of the elementary jumps between two consecutive 
active sites which displays a very wide distribution, and the temporal correlations
in those jumps.  There are two limit cases which have been thoroughly explored,
and which may serve as guides in the analysis:  one case focuses on temporal 
correlations with a narrow (say Gaussian) distribution of elementary jumps,
and the absence of a typical time scale would induce that the activity 
can be described as a self-affine profile in time, with a roughness 
exponent $1/z$.  The other limit case corresponds to the absence of 
temporal correlations focusing on the power-law distribution of elementary 
jump, and where the space-time map of activity can be seen as a Levy walk.
We do not know of any theoretical attempt to combine these two aspects to
get a general picture.  Due to the nature of the elementary jump
distribution, it is natural to explore the second limit case as a reference,
and check its domain of validity.
Thus we now consider the crude hypothesis that temporal
correlations can be neglected.
The probability $p(r,\tau)$ can thus be 
written as the convolution of $p(r,1)$ with itself 
$\tau$ times.  For large $\tau$, the distribution of $p(r,\tau)$
will therefore converge to a statistical distribution which is stable
for addition.  
If the distribution $p(r,1)$ has a finite second moment, then 
$p(r,\tau)$ should converge to a Gaussian law as a result of the central limit 
theorem.  However, this is 
never the case, since the above scaling implies that this would be obtained for 
$\alpha>3$, and we have seen that this case is similar to $\alpha=3$.  
In the intermediate case where $1\le \alpha \le 3$, we have seen above 
that $p(r,\tau)$ decays as $r^{-b}\propto r^{-\alpha}$, and hence, the second moment of
$r$ would diverge in an infinite system size.  As a result, the central 
limit theorem does not apply, but rather $p(r,\tau)$ converges to a stable
Levy law characterized by a power-law decay as $r^{-\alpha}$ for all 
$\tau$.  Hence, the power-law tail will be preserved, for large distances, 
i.e. as long as the correlation length is smaller than the system size, 
$\xi<L$.  This is indeed what is observed for $\alpha>1$.  
Still in the case where we neglect time correlations, we can relate the 
dynamic exponent $z$ to the large distance power-law decay through
\be
z=b-1=\alpha-1
\ee
We see from Figure~(\ref{fi:bzalpha}) that away from the value $\alpha=3$,
where $z\approx 2$, the above relation is poorly satisfied.  This 
indicates that temporal correlations becomes more and more important 
as $\alpha$ decreases, and cannot be neglected.  

When $\alpha$ approaches 1, all moments of the distribution become
controlled by the system size, and hence after a few time steps, 
$p(r,\Delta t)$ is smeared out over the entire domain size and 
no more power-law tail survives.  The scaling given by Eq.(\ref{eq:fur})
indeed breaks down for $\alpha\le 1$.  As a natural consequence, time 
correlations also vanish, and we enter a Mean-Field (MF) regime, which is 
independent of $\alpha$.

From these arguments, we arrive at a classification of different regimes
depending on the value of $\alpha$:\\
$\bullet$ $\alpha \ge 3$ the Laplacian regime, or the Edwards-Wilkinson 
regime with a quenched noise, where the kernel is equivalent to a second derivative.\\
$\bullet$ $1<\alpha < 3$ an intermediate regime, where the critical exponents
which are measured depend continuously on $\alpha$. \\
$\bullet$ $\alpha \le 1$ the Mean-Field regime, where spatial correlations are 
lost in a few time steps, and where --- as in the L-regime --- the value of 
$\alpha$ does not influence the critical exponents.

\subsection{Activity recurrence}

In the critical steady state, the activity map is highly correlated
and exhibits scale invariant features in both time and space. 
We studied previously the spatial distribution of activity after 
a fixed time lag $\Delta t$.  We now turn to a complementary description, 
focusing on a single site as a function of time.  More precisely,
if site $i$ was active at time $t_o$, we study the distribution, 
$p_{FIRST}(t)$, of the time delay $t$, such that the next 
move at site $i$ occurs at time $t_o+t$. Fig.~(\ref{fi:ptina}) shows such
a distribution for $\alpha=1,1.5,2,3$ and $\infty$. The very early time
behavior is dependent
on the distribution of displacement $u$ which is chosen.  If the
displacement is distributed over the interval $[a,b]$, for large $a$,
we prevent the recurrence of activity immediately after a move.  However,
this effect lasts only for a time much lower than the maximum time $T^*$ when 
$\alpha>1$. It does not affect the distribution over a large interval depending on the system size,
and thus can be discarded
from the analysis.  Then for $\alpha>1$, we observe a power-law decay 
\be
p_{FIRST}(t)\propto t^{-\tau_{FIRST}}
\ee
where $\tau_{FIRST}<2$. This power-law terminates at an upper cut-off 
which scales as the time necessary for a cluster to span the entire 
system size, $T^*=L^z$.  Let us now summarize our observations for various
$\alpha$ values.\\
$\bullet$ $\alpha>3$ Consistently with the previous discussion, 
the value of the exponent $\tau_{FIRST}$ is independent of $\alpha$ and 
amounts to $\tau_{FIRST}\approx 1.5 $ in good agreement with previous 
studies.
\\
$\bullet$ $2<\alpha<3$ The distribution drops rapidly (faster than any power-law)
above the cut-off scale $T^*$.  The exponent $\tau_{FIRST}$
progressively  increases and reaches the value $1$ as $\alpha$ tends to 2.
\\
$\bullet$ $1<\alpha<2$ The exponent $\tau_{FIRST}$ saturates to 1, for times
up to $T^*$, but in contrast to the previous cases, the distribution then
reaches a plateau from $T^*$ to $T^{**}$, before dropping faster than any
power-law.  The scaling of this second time scale (Fig.~\ref{fi:scaling}) is
identical to that of 
$T^*$,  $T^{**}\propto L^z$.  The relative importance of the $1/t$ region
and of the plateau, can be estimated by the ratio $T^*/T^{**}$ which goes
from 1 (no plateau) to 0 (no power-law) as $\alpha$ decreases from 2 to 1.
\\
$\bullet$ $\alpha<1$  We enter the Mean-Field regime where $p_{FIRST}$ is
constant for times up to $T^{**}$ which is proportional to the system size $L$.
Again this is consistent with the mean-field regime previously discussed.

It is interesting to note that the intermediate regime is now split in two 
cases, with a short and long time behavior differing for $1<\alpha<2$ rather
than a continuous evolution with an exponent $\tau_{FIRST}$ going from 1 to 0.
The flat plateau regime observed for $T^*<t<T^{**}$ and $1<\alpha<2$ is similar
to the Mean-Field regime for $\alpha<1$, and thus the transition 
to the MF regime appears not to be as brutal at $\alpha=1$ as proposed
from the scaling of the distance, but rather it turns out to be gradual, with
a mixed regime displaying non-trivial correlation for a ``macroscopic'' time $T^*$,
before reaching a Mean-Field behavior at later times.  The surprising
feature is that space and time correlations
do not disappear simultaneously. 
It might be interesting to note that the {\it mean} distance between
active sites separated by a delay $\Delta t$ also changes for $\alpha$
smaller or larger than 2. For $\alpha >2$, it follows from the
previous spatio-temporal analysis (part A), that the mean distance
between active sites scales as the correlation length $\xi$
since $b>2$. On the contrary, for $\alpha <2$, it scales as the system
size $L$. 

Let us consider an active site $i$ at time $t_0$. 
After a time $t$, the number of times, $n(t)$, the site $i$ has been active 
can be estimated as the total number of individual step proportional to $t$
divided by the number of sites where most of the activity is concentrated,
i.e. $t^{1/z}$, or 
\be\label{eq:n1}
n(t)\propto t^{1-1/z}
\ee  
Let us now estimate the same 
number using the distribution of activity recurrence $T$. From the power-law 
distribution of $T$, we can estimate the maximum among $n(t)$ numbers ---
supposed to be uncorrelated --- from 
\be
\int_{T_{max}(n)}^\infty p_{FIRST}(T)~dT \propto {1\over n(t)}
\ee
or
$T_{max}(n)\propto 
n^{1/(\tau_{FIRST}-1)}$.  Using the fact that $\tau_{FIRST}$ is always 
smaller than 2, we can compute the mean time between activity recurrence
$\langle T\rangle\propto T_{max}^{2-\tau_{FIRST}}\propto 
n^{(2-\tau_{FIRST})/(\tau_{FIRST}-1)}$.  This same average time is 
also equal to $\langle T\rangle =t/n(t)$.  Equating these two estimates
gives
\be\label{eq:n2}
n(t)\propto t^{\tau_{FIRST}-1}
\ee
Comparing Eqs.~(\ref{eq:n1}) and (\ref{eq:n2}), we arrive at the 
following expression for $\tau_{FIRST}$:
\be\label{eq:tauz}
\tau_{FIRST}={2z-1 \over z}
\ee
This estimate is based on the fragile assumption that time correlations 
can be neglected.  We have seen above that a similar hypothesis lead to
$z=\alpha-1$, from what we deduce
\be
\tau_{FIRST}={2\alpha-3\over \alpha-1}
\ee
This expression however requires that the distribution of $T$ is 
not governed by the upper cut-off $L^z$, and thus $\tau_{FIRST}>1$ strictly,
or $\alpha > 2$ in the above formula.  Figure~(\ref{fi:taualpha}) shows the
measured value of $\tau_{FIRST}$ as a function of $\alpha$, together 
with the above relation.  We observe a reasonnable agreement for $2<\alpha<3$. 

\subsection{Avalanche Dynamic}

Up to now, we have considered an ideal driving where the front 
displacement is controlled by adjusting instantaneously the driving
force.  Assuming an overdamped dynamics, we can reconstruct a different
type of driving.  

In order to proceed, it is useful to introduce the notion of avalanches
or bursts.  Let us consider the sequence of external force $F_{ext}(t)$.
Any force $F$ will split the signal into consecutive intervals 
with alternating $F<F_{ext}$ domains (or ``obstacles''), and $F>F_{ext}$
intervals (or ``bursts'').  Each  burst is thus the front motion
which would result from a constant force $F$ being imposed to the 
system.  

Avalanches were introduced initially\cite{RGuyon} in order 
to understand some features of invasion percolation\cite{Furuberg}.  However, 
in Ref.\cite{RGuyon}, some approximations were unjustified or unfounded.
A complete solution of the avalanche statistics applied to
invasion percolation and other extremal models is presented in 
Paczuski et al\cite{Maslov}.  The key property is that it is 
possible to relate the avalanches in the external forcing to the
cluster statistics in some extremal models such as invasion percolation.  
The avalanche size distribution at fixed $F$ is a power-law distribution 
up to a maximum avalanche size which diverges as $F$ approaches to
a critical value $F_c$.  This mapping allows to derive geometrical
information on the activity from a simply accessible external signal, 
the driving force.  One amazing feature is that the latter is one-dimensional 
but encode a multi-dimensional information.

In our case, it is 
unfortunately impossible to establish such a direct mapping because of 
the non-local nature of the interaction.  Hence, the connectedness of the 
clusters is lost, or equivalently, the clusters to be defined from the 
driving force avalanches do not have a straightforward geometrical 
interpretation.  Nevertheless, the maximum external force $F_c$ encountered in the 
external driving can still be identified with a critical point where the 
correlation length diverges.  This critical point is a depinning transition.
For a constant driving $F$ strictly smaller than $F_c$, the front will 
advance only over a finite distance and stop when it encounters a pinning force
larger than $F$.  Similarly, if $F>F_c$, the front will never be stopped.

It is also possible to use the time evolution of the system to define
two distributions of avalanche which do
not require the precise identification of the critical threshold. At
each time step, $t$, one can construct the avalanches which correspond
precisely to the loading $F(t)$. In fact, two such avalanches can be
considered, for times either larger or slower than $t$, the {\it
  forward} and {\it backward} avalanches respectively as shown in Figure
\ref{fi:fct}. Figure \ref{fi:ava} 
shows both cumulative avalanche distributions for our model with
$\alpha >1$. They behave as power laws 
\be
N_{f,b}(t)\propto t^{1-\tau_{f,b}}
\ee
with no upper cut-off apart from the system size.

Maslov\cite{Maslov_alone} showed that for a variety of models, 
the exponent $\tau_f$ 
was super-universal, and equal to 2. Consequently, forward avalanches 
do not reveal much information on the model.  However, in the extremal 
models studied in Paczuski et al\cite{Maslov}, the backward avalanche 
exponent is not super-universal and it does give rise to a non trivial
critical exponent $\tau_f$.

In our numerical results, the 
universal result $\tau_f=2$ is observed for $\alpha > 1$. 
The backward value is also measured to be $\tau_b=2$ for the same values of $\alpha$. 

Such power-laws with exponents $\tau_f=\tau_b=2$ also appear in a series of
uncorrelated random numbers picked from the same distribution,
however, we cannot infer from this result that the forces are
uncorrelated. They are indeed strongly correlated as can be seen from
the power spectrum of the signal $F_{ext}(t)$ (Fig.\ref{fi:fspec}) which
shows a power-law extending over long time intervals (yet smaller 
than the time needed to span the system $L^z$).  If we were to interpret 
this signal as self-affine we would estimate from such a graph a self-affine
Hurst exponent $\zeta_F=-0.4$, a negative exponent which means that the signal 
is strongly anti-correlated.   The same power-law appears for all values of 
$\alpha$ larger than 1.  One has to be cautious while manipulating 
Hurst exponent which are negative, since most properties relying on 
the self-affine nature of a signal break down in this case. In fact  
for most properties, such functions behave as if the Hurst exponent
was actually zero (white noise).  (Otherwise, in order to 
reach reliable conclusions, one may consider the time-integrated force signal
i.e. the energy dissipated during the front motion.   More precisely, the mean 
force can be substracted from $F$ before integration so that only the
fluctuating 
part of the energy is considered.  For $\alpha >1$, this signal has a self-affine 
exponent $\zeta_F+1\approx 0.6$, which can be measured using standard tools.)
The avalanches are thus expected to behave as if the force signal was a 
white noise, hence $\tau_f=\tau_b=2$. For small exponent $\alpha <1$, there is a short time 
(high frequency) behavior which seems to follow a similar power-law behavior.
The plateau appearing for very small times is probably due only to numerical 
accuracy and has not to be considered.  
But at large times, the power spectrum of the force crosses over to a 
$\omega^{-2}$ regime which is the signature of a $\zeta_F=0.5$ self-affine 
signal, analogous to a random walk. The avalanche exponents are consequently 
changed to $\tau_{f,b}=1.5 \pm 0.05$ for $\alpha<1$.

Although the previous analysis of avalanches can in principle be accessed
experimentally, it requires to be able to switch from a 
displacement-controlled to a force-controlled mode, what is in general 
a rather difficult requirement.
There is another, more natural, way to have access to the fluctuating 
driving force through the intermittent front motion.  We have seen that
a constant force control for  finite size system either lead to a stop
or a constant motion, the transition between the two regime corresponding 
to the maximum $F_{ext}$.  Thus a single configuration will determine the 
threshold, and for a finite size system, this single point generally is not
representative of the critical behavior, and it contains a rather poor 
information.  A way to circumvent this difficulty is to introduce a small
but non-zero stiffness $\epsilon$ in the control (Fig.\ref{fi:fel}), so
that as the front 
advances, the driving force is progressively decreased by a quantity  
$\epsilon. \langle u\rangle$.  This insures that the motion will never 
be unlimited.  As soon as the front is pinned, then the driving force
is slowly increased up to the depinning limit.  If $\epsilon$ is small 
enough, only the forces close to
the pinning threshold will be probed.  In addition, the statistics of 
the front advance between two pinned configuration gives some information 
on either the correlations of $F_{ext}$, or on its statistical distribution 
close to the critical threshold.

We performed such an analysis for various values of $\alpha$.
We observed that the avalanches have a maximum size $s^*$ 
which depends on $\epsilon$ as 
\be
s^*\propto \epsilon^{-\kappa}
\ee
with 
\be
\label{eq:kap}
\kappa\approx 0.65
\ee
for all values of $\alpha$ investigated from $1$ to $4$.  This 
maximum size can be used to scale the avalanche size $s$
(Fig.\ref{fi:fspec}), and thus it 
allows to account for the distribution in an $\epsilon$ independent manner
\be
n_\epsilon(s)\propto {1\over s^{*r}}
\psi\left({s\over s^*}\right)
\ee
where $\psi$ is a power-law distribution for small arguments $\psi(x)\propto
x^{-\theta}$ for $x\ll 1$, and decays rapidly to zero for $x\gg 1$.
And $s^{*r}$ is a normalization factor.

There are different features which may contribute to the statistics of 
$s$.  We  here focus on one aspect of the problem namely the statistical 
distribution of the pinning force close to the depinning critical point.
Note that the results which could be derived from the self-affine
structure of the signal $F_{ext}(t)$ are not valid due to the
asymmetry of the distribution of forces, $p(F_{ext})$ (Fig.\ref{fi:fc}).

Let us note 
\begin{equation}
p(F_{ext})=(F_c-F_{ext})^\beta\equiv\delta^\beta
\end{equation}
the distribution of external forces near the threshold $F_c$. 
After $n$ events, the system has lost an elastic force $\delta$, 
$$
\delta=n\epsilon. 
$$
The maximum force encountered follows
$$
\delta^{-(1+\beta)}=n
$$
Balancing the two expressions leads to
$$
\epsilon=n^{-1/(1+\beta)-1}=n^{-(2+\beta)/(1+\beta)}
$$
hence
\be
\kappa={1+\beta\over 2+\beta}
\ee
From numerical data, we have $\beta\approx 1$. It leads to
$$\kappa\approx 0.66$$ 
in good agreement with Eq.~(\ref{eq:kap}), although temporal correlations have 
not been considered here.    
  
\section{Discussion}
We have presented in this article a study on the correlations induced
by the dampened motion of a finite size elastic chain with long range
interactions and driven along its transversal direction on a surface with 
quenched disorder.

We have shown that the system organizes after a transient in a stationary state 
with long-range correlations. The memory of the initial state is lost. The final 
state does not depend on the initial state, but its characteristics depend on 
$\alpha$. We have pointed out three regimes: For very long range couplings 
($\alpha\le D$), a regime controlled by finite size effects and analogous to a 
Mean-Field one. For very short range couplings ($\alpha \ge D+2$), a regime 
controlled by small distance singularities of the Green function and displaying 
the same characteristics as Laplacian couplings.
In between, ($D<\alpha< D+2$), an intermediate regime where characteristic 
exponents $b$, $z$ and $\zeta$ evolve continuously with $\alpha$.
The transition between the three regimes of the activity map is a new one. It 
is summarized in figure \ref{fi:bilan}. The transition to a Mean Field uniform 
regime does not appear simply for $\alpha=3D/2$ as been proposed by 
D.Fisher et al.\cite{Fisher} by Functional Renormalization Group analysis, but it 
appears clearly from the simulations, that spatial and temporal correlations do 
not disappear simultaneously. The transition to the Mean Field regime is 
characterized by a cross-over time-length $t_c$ growing with $\alpha$,
above which the
uniform distribution is valid and below which the power law behavior
with non trivial exponent is observed. Even if the strong pinning
assumption ensures that the instabilities are local, when $t_c$
reaches small times, the behavior of the system becomes
``delocalized''. The ``delocalized'' behavior
refers to a uniform spreading of activity over the system above $t_c$. 
For $\alpha >2$, the system exhibits ``collective'' organization 
(characterized by long-range correlations) but ``localized''
behavior. For $\alpha <1$ the system exhibits a uniform
``delocalized'' behavior analogous to the Mean Field one. 
For $1<\alpha<2$ the system exhibits collective organization
with distortions at small scales but ``delocalized'' behavior at large
scales (above $t_c$). Moreover, the time
$t_c$ goes to infinity in the thermodynamic limit: it scales as the activity 
spreading time $L^z$ and does not introduce a new scaling. Its origin remains 
unclear to us. An analogous transition appears in the external force for 
$\alpha <1$ (see Figure\ref{fi:fspec}).

Modifying the parameters of the system, combination of the above behavior 
can also be found.
For a mix of various couplings for example, the system exhibits a
cross-over between a regime controlled by the short range coupling (at
small time and small length) and a regime controlled by the long range
coupling (at large time and large length). The transition between these
two regimes is size independent and the spatial and the temporal
transition are related by the dynamical exponent of the shortest
couplings. 

The results presented above have been obtained with a discretized system and 
periodic boundary conditions. We have checked that the same results are 
obtained, far away of the boundaries, without periodic boundary conditions. 
Particularly, the exponents obtained are the same in this case. The temporal 
discretization and the breakdown of the translational invariance (because only 
one site is allowed to jump at each time step) are not limiting. They are 
consistent with the ``strong pinning assumption''. At this stage, the simulations do
not allow us to justify this assumption more strongly. We know however
\cite{tangroux} that there exists a transition from weak to strong
pinning which could be used to justify a modeling at larger
scales. But for very long range (small $\alpha$) and small system
size we may encounter situation of weak pinning, a case out of reach
for our model. We suppose that the local curvature of the
pinning potential is always sufficiently high to maintain the system
in the strong pinning limit. In this limit, the results presented
here appear to be universal, provided some quenched disorder is introduced: the
choice made for the distribution $u$ and for the time step only affects 
short distances. As has been mentioned, they are important in the mean
field regime, in agreement with ref.\cite{Fisher} but they do not affect the
{\it position} of the transition to the Mean Field behavior. When no
disorder is 
introduced, the system evolves toward a limit cycle of period $L$. The
boundary conditions affect the distributions only at the proximity
of the upper cut-off scale, $L$ or $L^z$, and the displacement fields near the
borders. When disorder is present, it is possible from our
numerical results to
find intermediate scales where the behavior of the system is
characterized in space and in time by power law distributions. The
critical exponents of the distributions only depend on $\alpha$ as discussed 
above.
More surprisingly, due to the temporal anticorrelation in the
threshold force, the internal avalanche distribution seems to be independent of
$\alpha$ and does not reflect the correlations of the external
force. A better accuracy is restored with the help of an elastic external
driving. This gives rise to external avalanches.
Let us mention that velocity has to be defined carefully in the case of driving 
with constant external force\cite{Raphael}.

Let us turn to the notion of multistability. For a one particle
system, the multistability is due to the non
linear concurrent effect of the elastic force and of the pinning
force. The condition of
multistability can be given by a linear stability analysis. The particle
is unstable when the negative curvature of the pinning potential
overcomes the elastic stiffness. It leads to
hysteretic behavior \cite{Caroli}. The energy dissipated at very low
velocity is
related to the area of the hysteretic cycle. For a $N$-particle system,
the instability involves more than one particle, due to the
coupling between asperities. The size of the domain involved in the
instability depends on the distortions in the curvature of the pinning
potential and on the range of the elastic interactions. In the ``weak
pinning limit'', the one-particle instability condition is not
effective, but multistability can occur at larger scale. The minimal
size giving rise to multistability is usually computed by the ``Larkin
Method'' \cite{Larkin}. This length is associated to a dynamic by
blocks whose existence remains controversial in the literature.
In the ``strong pinning limit'', the one particle case
is effective,
but the elastic coupling may propagate the instability to an avalanche
of larger size. We showed here
that the system thus organizes into a ``critical state''. There is no
intrinsic length scale in such states, but the lowest
cut-off and the system size. This effect
gives rise to ``kinetic roughening'' in finite systems and has been
well studied in the case of short range elastic interactions for
various type of pinning \cite{cont}. It does not induce block motion
involving more than one asperity. An interesting question
would be: is the ``block dynamics'' of the weak pinning limit analogous
to the strong pinning dynamics? This question remains open. 

\section{Conclusion}

We have shown in this paper that in the strong pinning limit and for a
quasi-static driving, a finite
elastic system driven on a disordered surface exhibits long range
dynamical correlations.
This can be shown by the roughening of the surface, in agreement with
spreading or crack front propagation experiments. It can be studied
numerically by the statistical analysis of the Activity Map.
The range $\alpha$ of the interaction allows
to interpolate between a Laplacian dynamical regime (for $\alpha
>\alpha_{c2}$) and a Mean Field dynamical regime (for
$\alpha<\alpha_{c1}$). In our one-dimensional simulations, with quenched
decorrelated disorder, $\alpha_{c1}=1 (=D)$ and
$\alpha_{c2}=3 (=D+2)$. The transition to the Mean Field regime appears to take place
progressively from the large times when $\alpha<2$. Space and time
correlations do not disappear simultaneously. 

The internal avalanche size distribution seems to be ``superuniversal''
(independent on $\alpha$) because it
does not preserve informations on the temporal correlations in the
threshold force and exhibits the simple $1/f^2$ behavior
characteristic of jumps. The dependence on $\alpha$ is recovered by an 
elastic external loading. 

This study shows one example of physical systems where microscopic
disorder leads to large scale structures as can be revealed by kinetic
roughening (large scale displacement fluctuations) or avalanche
distributions. The algebraic decay of the
interactions determines the critical exponents but other details are 
unimportant  in the scaling behavior. The large scale power-law behavior is 
a signature of the self-organization of the system. 

This study was designed to describe a ``strong pinning'' limit. It would be
of interest to extend the model to the weak pinning case.

\acknowledgements

It is a pleasure to acknowledge fruitful discussions with 
J. Schmittbuhl, B. Protas, S. Krishnamurthy, H.J. Herrmann 
and M. Fermigier. 
This work is partly supported by the Groupement de Recherche 
``Physique des Milieux H\'et\'erog\`enes Complexes'' of the CNRS.

%%%%%%%%%%%%%%%%%%%%%%%%%%%%%%%%%%%%%%%%%%%%%%%%%%%%%%

\end{multicols}
%%%%%%%%%%%%%%%%%%%%%%%%%%%%%%%%%%%%%%%%%%%%%%%%%%%%%%
%%%%%%%%%%%%%%%%%%%%%%%%%%%%%%%%%%%%%%%%%%%%%%%%%%%%%%
List of Captions:

\begin{figure}
\caption{\label{fi:des} 
(a) Schematic motion of the elastic chain in the force field $\eta(x,u(x))$
between $t$ (black) and $t+\Delta t$ (grey). The redistribution of the elastic forces
via the long-range interactions is not visualized in this picture.
(b) Schematic description of the pinning force for one site $x$. The abscisse is 
now the average position of the line.}
\end{figure}
\begin{figure}
\caption{\label{fi:acti} 
Activity map showing the location of the active sites as a funtion of
time for various couplings $\alpha=1 $(a), $\alpha=2 $(b) and $\alpha=3
$(c). The system size is L=1024.}
\end{figure}
\begin{figure}
\caption{\label{fi:depla} 
Displacement front at different time intervals and for various
couplings $\alpha =1 $(a),$\alpha =2 $(b) and $\alpha =3$(c). The system
size is L=1024. }
\end{figure}
\begin{figure}
\caption{\label{fi:rugo} 
Average power spectrum of the displacement front position in log-log
scale for a $\alpha=2$ coupling. The dotted line shows a power-law fit
which corresponds to the roughness exponent $\zeta=0.35$. The system
size is $L=1024$.}
\end{figure} 
\begin{figure}
\caption{\label{fi:zetalpha} 
Roughness exponent $\zeta$ versus coupling range $\alpha$, measured
from the power law fits of the displacement correlation function.}
\end{figure} 
\begin{figure}[hbt]
\caption{\label{fi:fc} 
(a) Statistical distribution $n(f)$ of forces along the chain in the stationary
regime, for various couplings $\alpha$. The scale for forces is given
by the maximum of forces $\vert G(x^*,x^*) . max(u)\vert$. The pinning threshold 
is taken arbitrarily as $\eta_{i(x^*)}(x^*)\equiv -1$. (b) Contribution
of the maximum
depinning forces $f(x^*)=1-F_{ext}(t)$ (little $\diamond$) in the
distribution $n(f)$ for $\alpha=1$. The dashed line stands for a fit
of $n(f(x^*))$ near the threshold $f(x^*)_c=1-F_c/L$ with $\beta=1$.}
\end{figure} 
\begin{figure}[hbt]
\caption{\label{fi:Furuberg} 
Data collapse of seven different probability distributions $p(r,\Delta
t)$ for $\alpha=2$ and time intervals ranging from 1 to 64. The system
size is L=512. The redistribution exponent is $b=1.95$, and the best
data collapse is obtained for $z=1.30$.}
\end{figure} 
\begin{figure}[hbt]
\caption{\label{fi:bzalpha} 
(a) Plot of the $b$ exponent. (b) Plot of the $z$ exponent obtained for
different values of $\alpha$. The dotted and dashed lines are the
proposed asymptotic behavior discussed in the text. The dotted line
($2\alpha/3$) on the right take account for the relation $z=1+\zeta$ and the
expression proposed for $\zeta$ by Fisher et al. } 
\end{figure} 
\begin{figure}[hbt]
\caption{\label{fi:ptina} 
Log-Log plot of the first return probability distribution
$p_{FIRST}(t)$ as a function of the time $t$ for various coupling
range $\alpha=1$, $\alpha=1.5$, $\alpha=2$, $\alpha=3$ and
laplacian. The system size is $L=512$. } 
\end{figure} 
\begin{figure}[hbt]
\caption{\label{fi:scaling} 
Data collapse of the first return probability distribution for various
system sizes for $\alpha=1.5$. The upper cut-off scale as $L^z$ as
well as the extent of the plateau.} 
\end{figure} 
\begin{figure}[hbt]
\caption{\label{fi:taualpha} 
Plot of the $\tau_{FIRST}$ exponent obtained for
different values of $\alpha$. Two different exponents appear for $1<\alpha<2$ 
in the Mixed Regime. They correspond to the distribution below and above $t_c$.
The dotted lines are the proposed asymptotic behavior discussed in the text. } 
\end{figure} 
\begin{figure}[hbt]
\caption{\label{fi:fct} 
External loading $F_{ext}(t)$ as a function of time. From this signal, on can define
avalanches either by prescribing a fixed value of $f$, and considering
the intervals where $f$ remains below this limit, or by constructing
these avalanches starting from all forcing values $F_{ext}(t)$ with the
statistics issued from the model, and choosing either a forward or
backward time direction.} 
\end{figure} 
\begin{figure}[hbt]
\caption{\label{fi:fel} 
Instantaneous loading $F_{ext}(t)$ as a function of time and external
elastic load of stiffness $\epsilon$.} 
\end{figure} 
\begin{figure}[hbt]
\caption{\label{fi:ava} 
Log-log plot of the distribution of the forward and backward
avalanches for $\alpha>1$. The straight lines are power-law fits of exponents $-2$.} 
\end{figure} 
\begin{figure}[hbt]
\caption{\label{fi:fspec} 
Spectral amplitude of the time fluctuation of the time integration of the
threshold force $F_{ext}(t)$ for various couplings $\alpha$ ranging from $0$ to
$3$. The straight line represents a power-law of exponent $-2.2$ which
implies a roughness exponent of the direct time force signal of
$-0.4$ for $\alpha > 1$. The full line represents a power-law of
exponent $-4$ which implies a roughness exponent $+0.5$.} 
\end{figure} 
\begin{figure}[hbt]
\caption{\label{fi:fava2} 
Log-log plot of the distribution of external avalanches for $\alpha=2$
and for ten different stiffnesses $\epsilon$ ranging from $10^{-4}$ to
$10^{-3}$. The straight line is a power-law fit of exponent
$-1.09$. The upper cut-off scales like $s^*\propto\epsilon^{-0.65}$.} 
\end{figure} 
\begin{figure}[hbt]
\caption{\label{fi:bilan} 
Schematic transition between the three apparent dynamical
regimes. Dotted line separates the classical power law regime and a new one
regime with a ``local'' behavior at short time and a ``delocalized''
behavior at large time. The Mixed Regime stands for a regime with
Mean-field type character only for time correlations and at large times.} 
\end{figure}

\begin{figure}[hbt]
\centerline{\epsfig{file=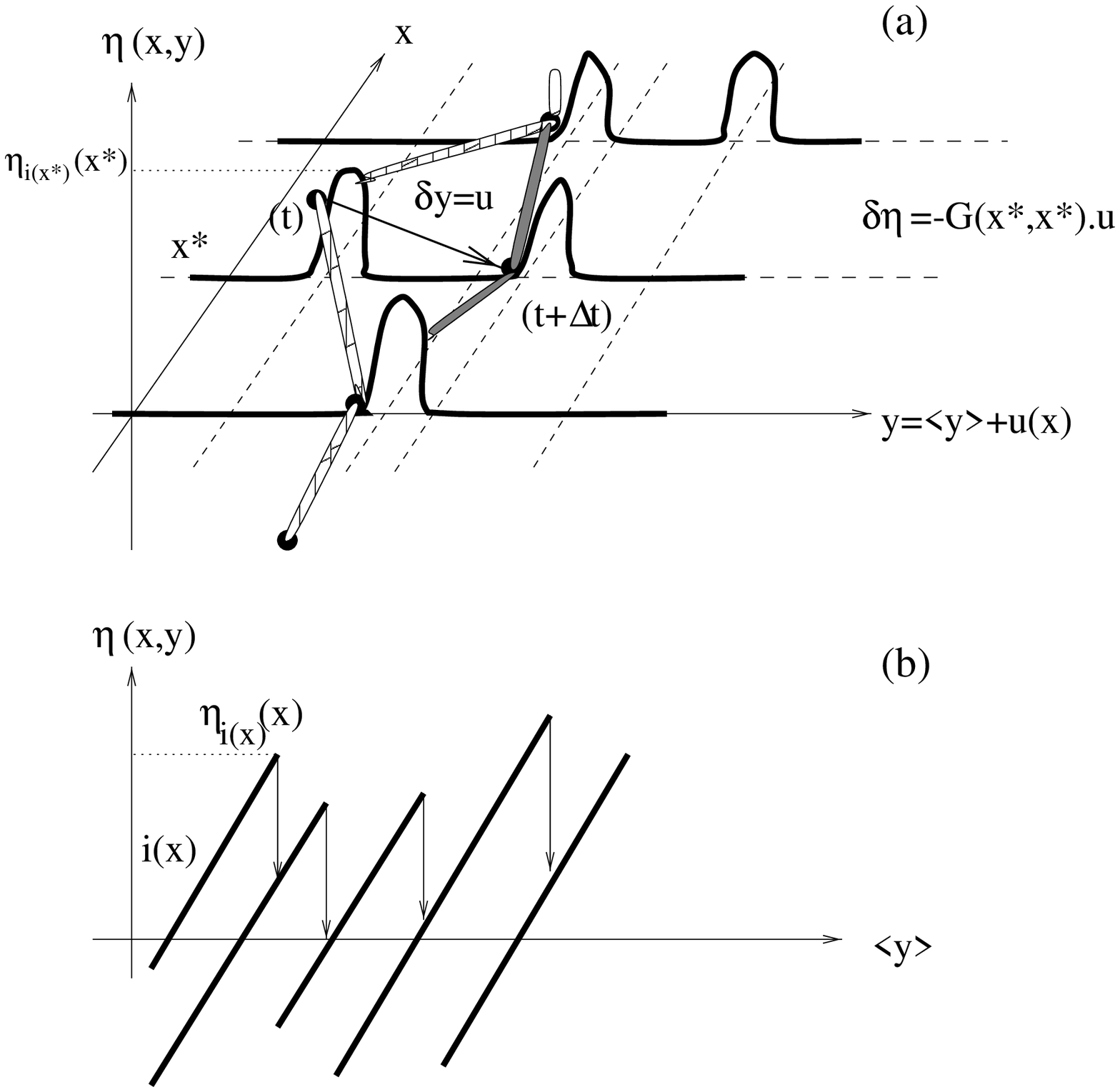,width=10cm,height=14cm,angle=0}}
\begin{center} Figure 1\end{center}
\end{figure}

\eject
\begin{figure}[hbt]
\centerline{\epsfig{file=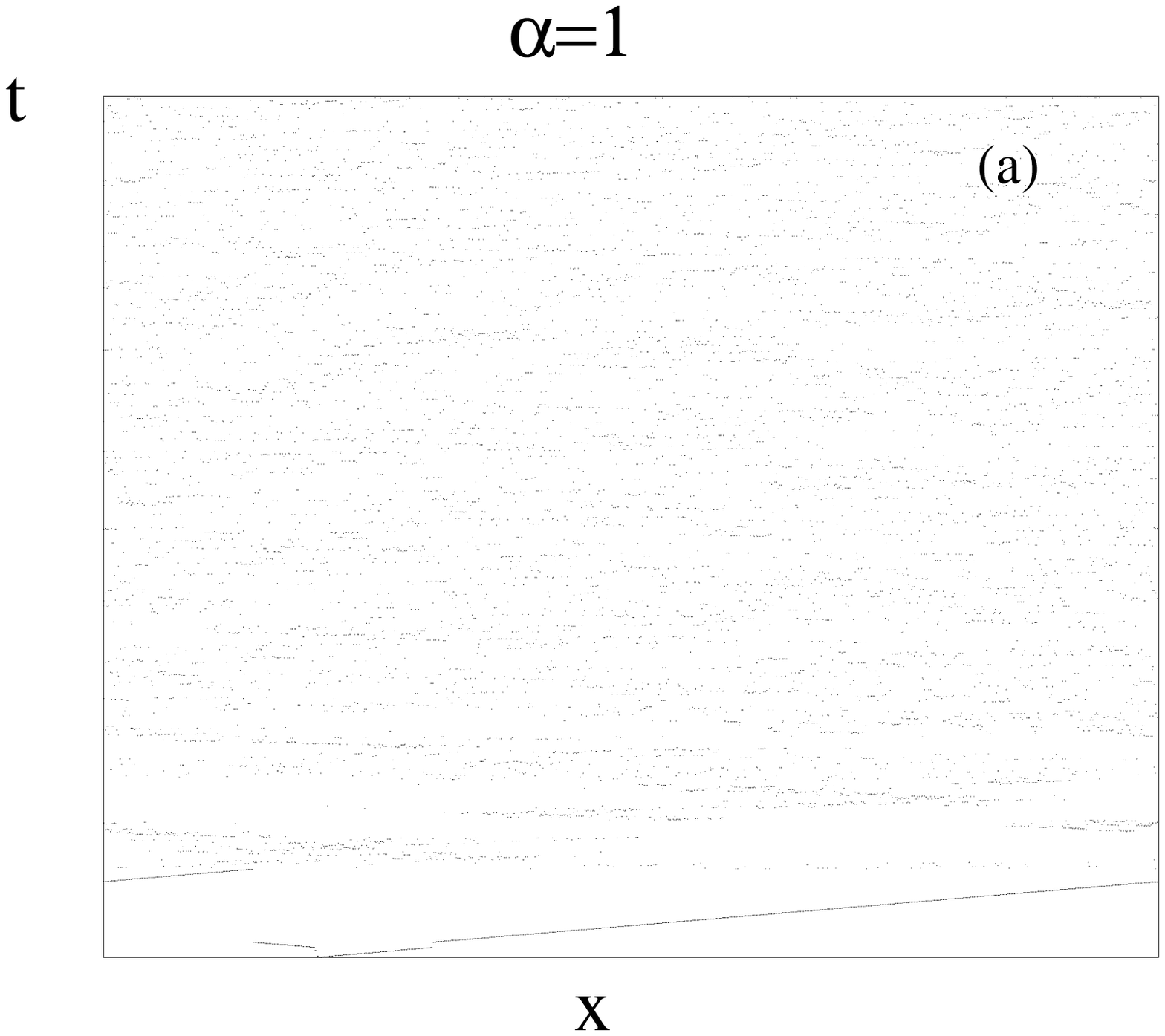,width=6cm,height=10cm}
            \epsfig{file=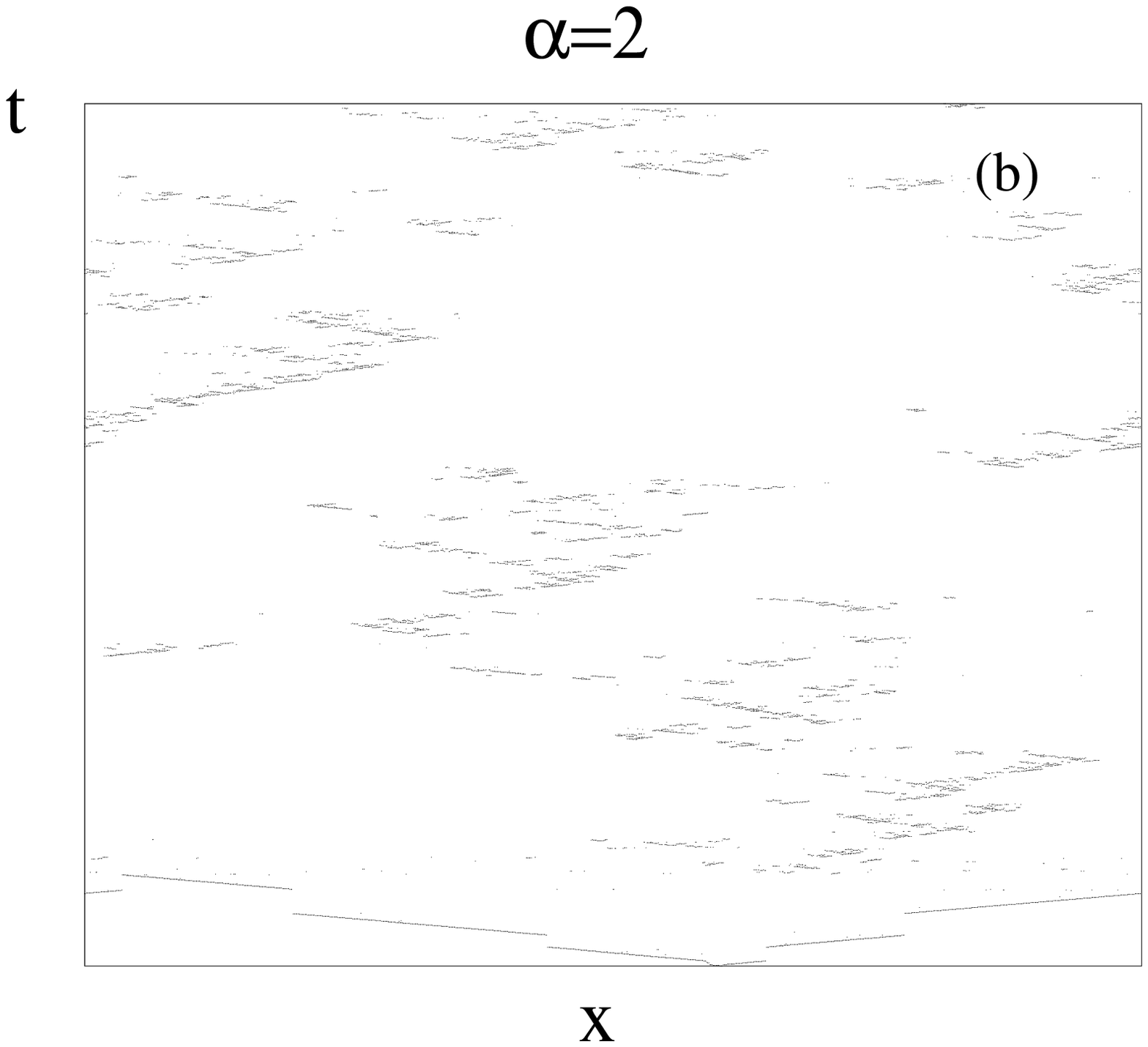,width=6cm,height=10cm}
            \epsfig{file=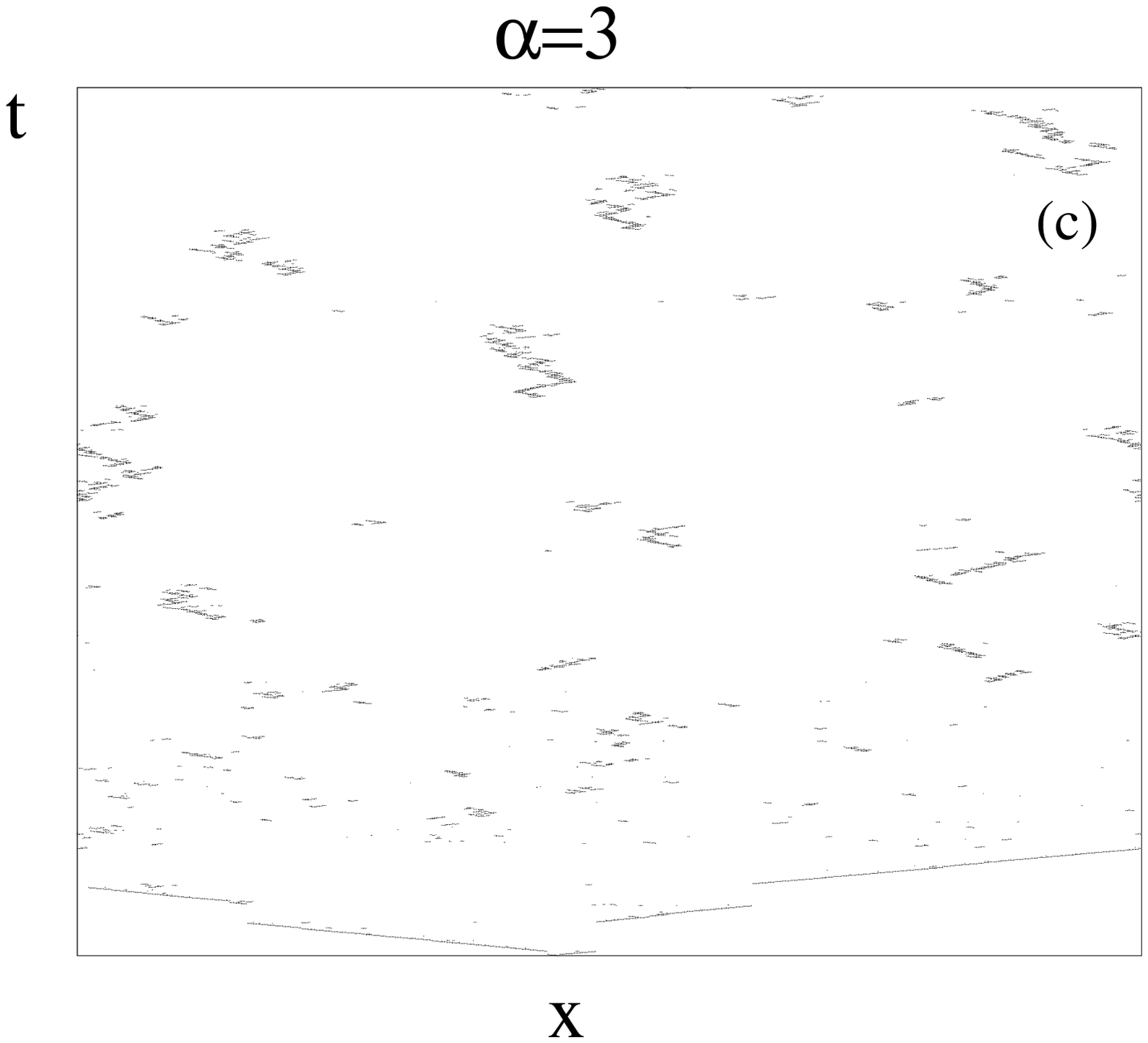,width=6cm,height=10cm}}
\begin{center}Figure 2\end{center}
\end{figure}
\begin{figure}[hbt]
\centerline{\epsfig{file=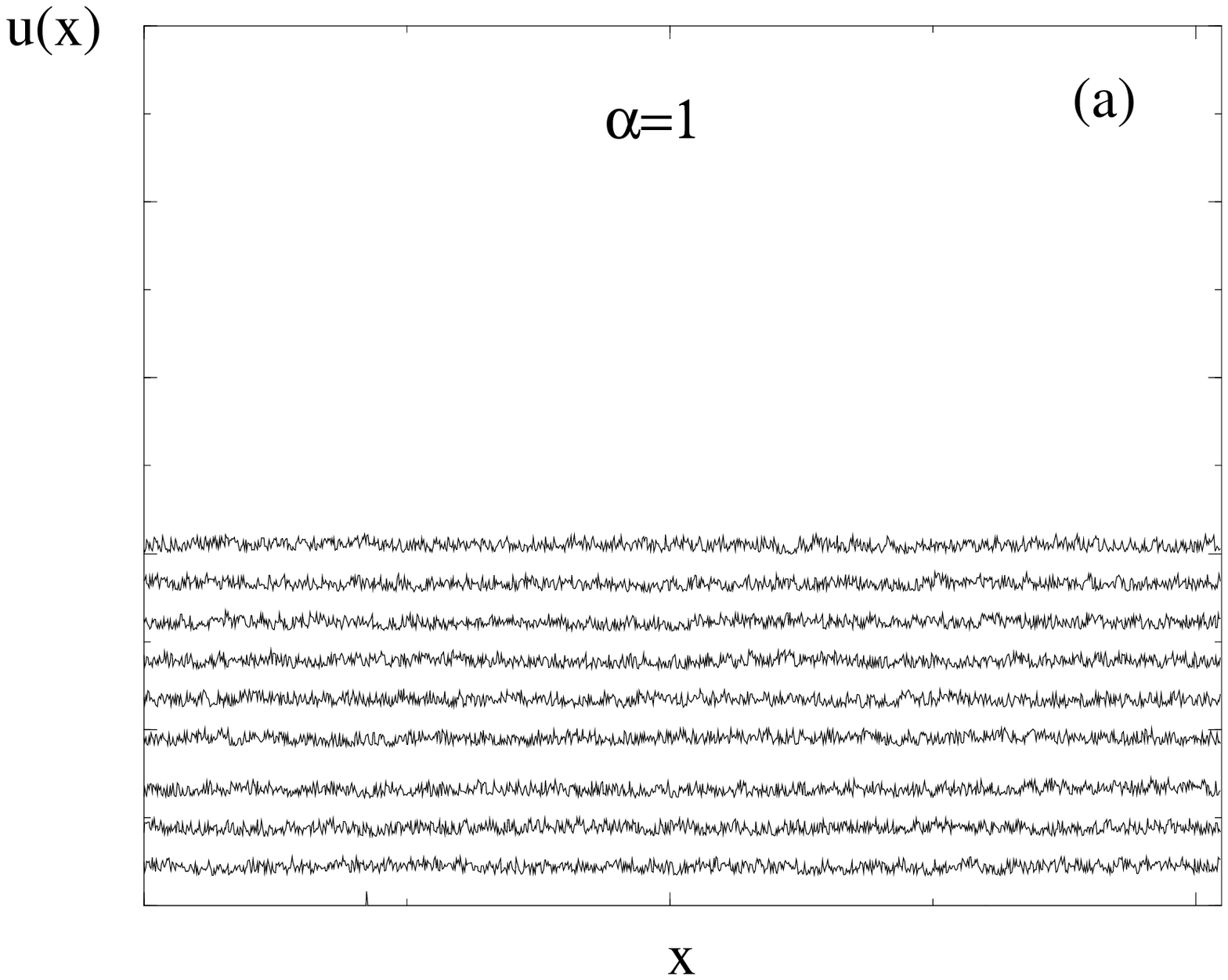,width=6cm,height=6cm}
            \epsfig{file=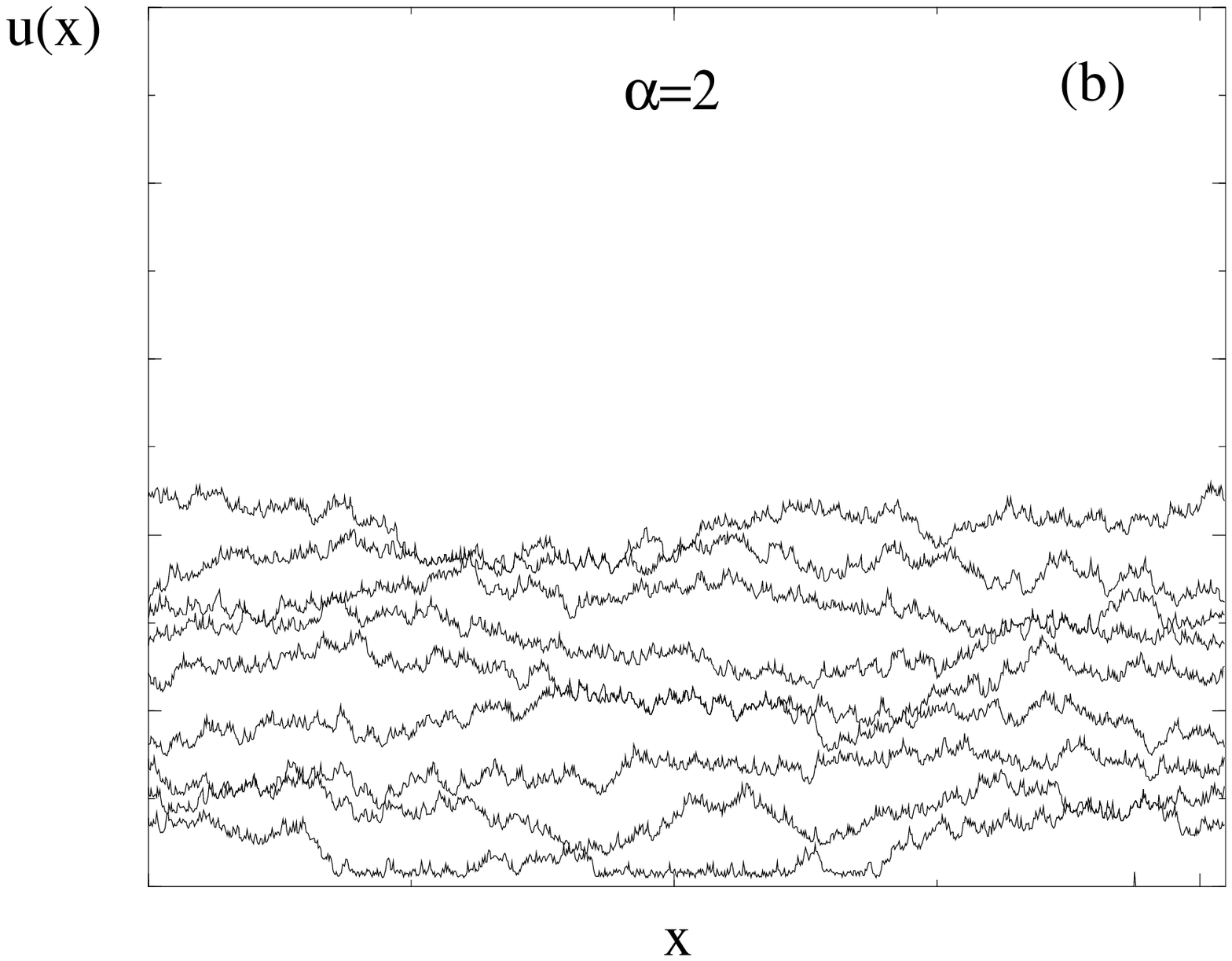,width=6cm,height=6cm}
            \epsfig{file=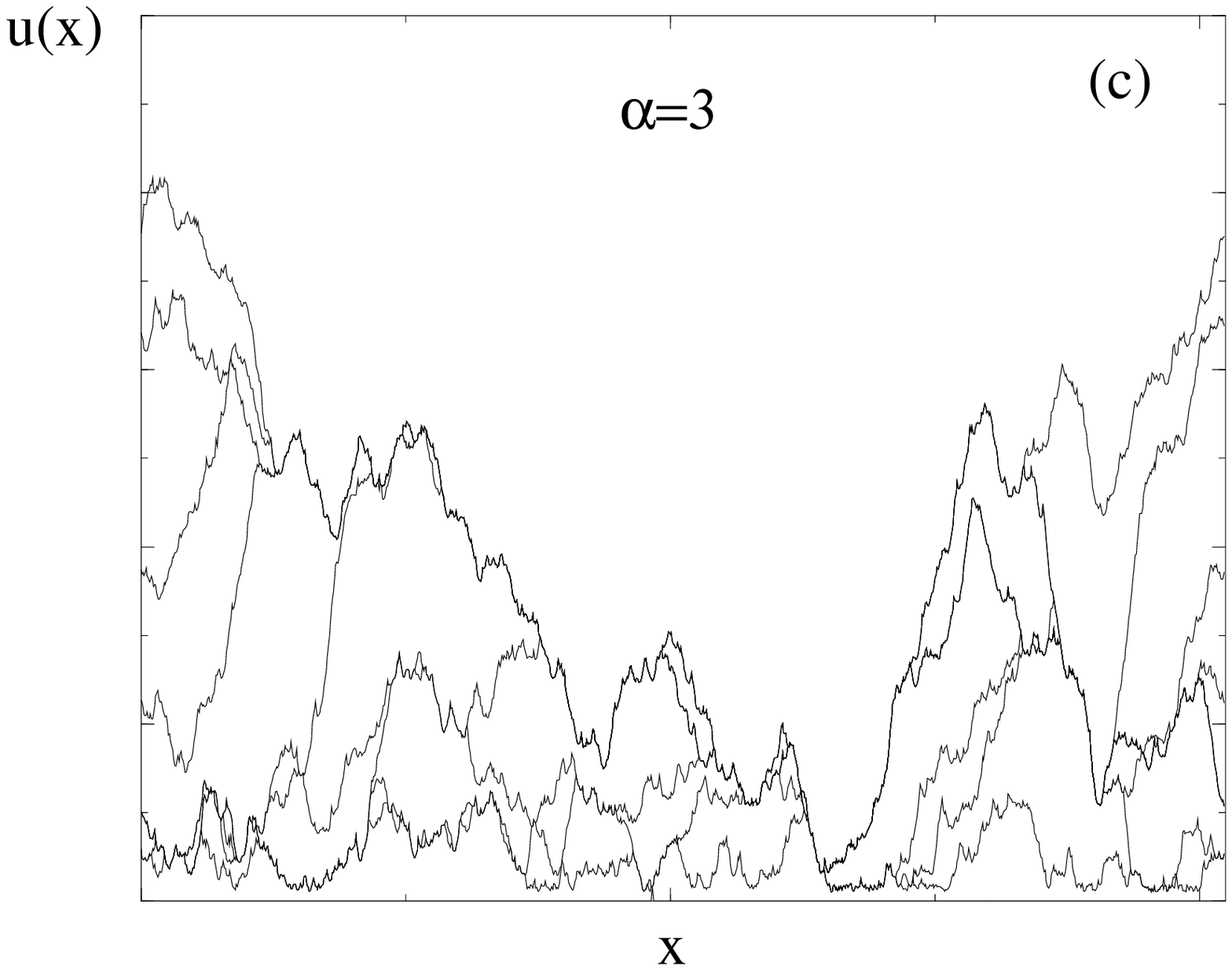,width=6cm,height=6cm}}
\begin{center}Figure 3\end{center}
\end{figure}

\eject 
\begin{figure}[hbt]
\epsfxsize=\columnwidth\epsfbox{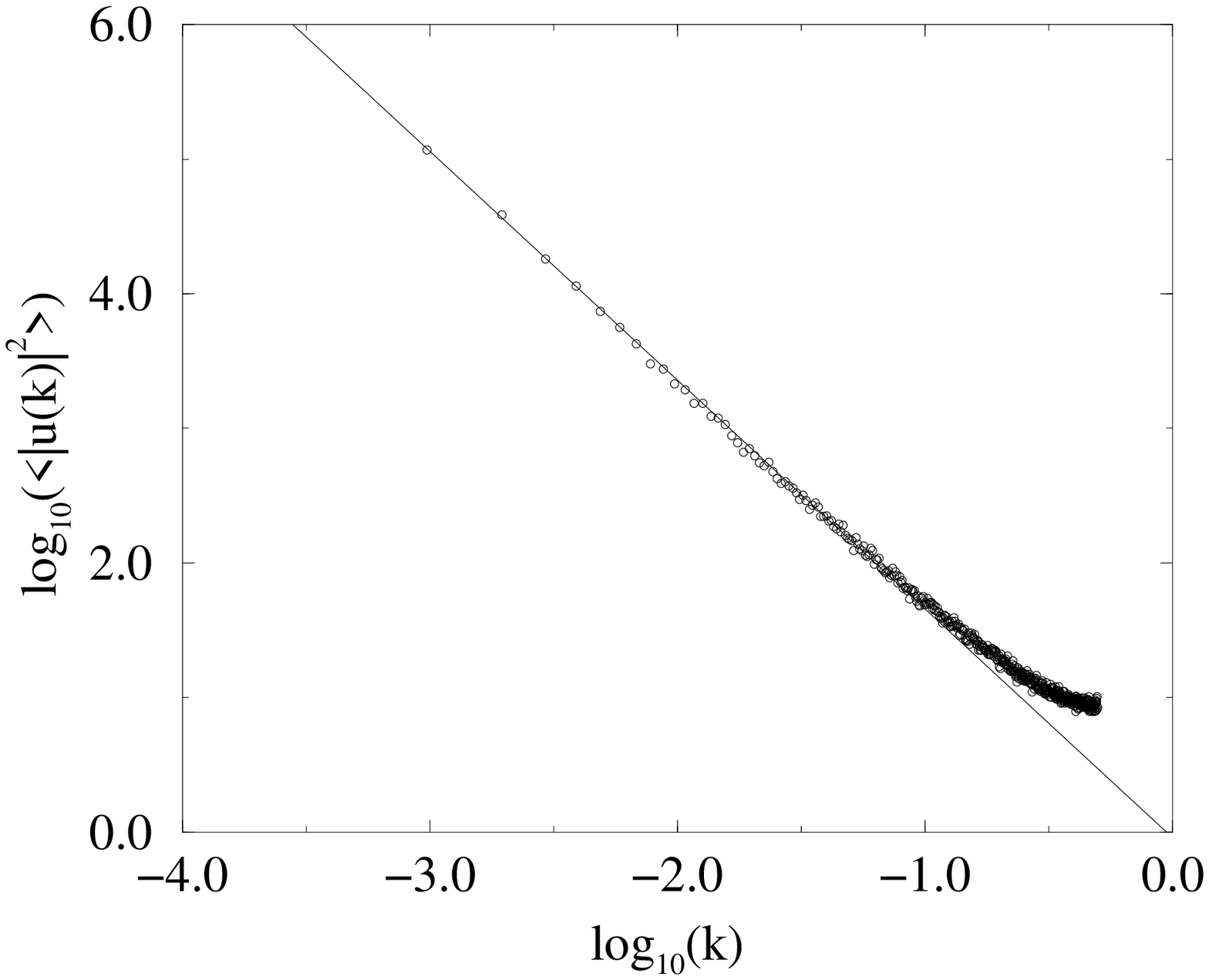}
\begin{center}Figure 4\end{center}
\end{figure} 
\begin{figure}[hbt]
\epsfxsize=\columnwidth\epsfbox{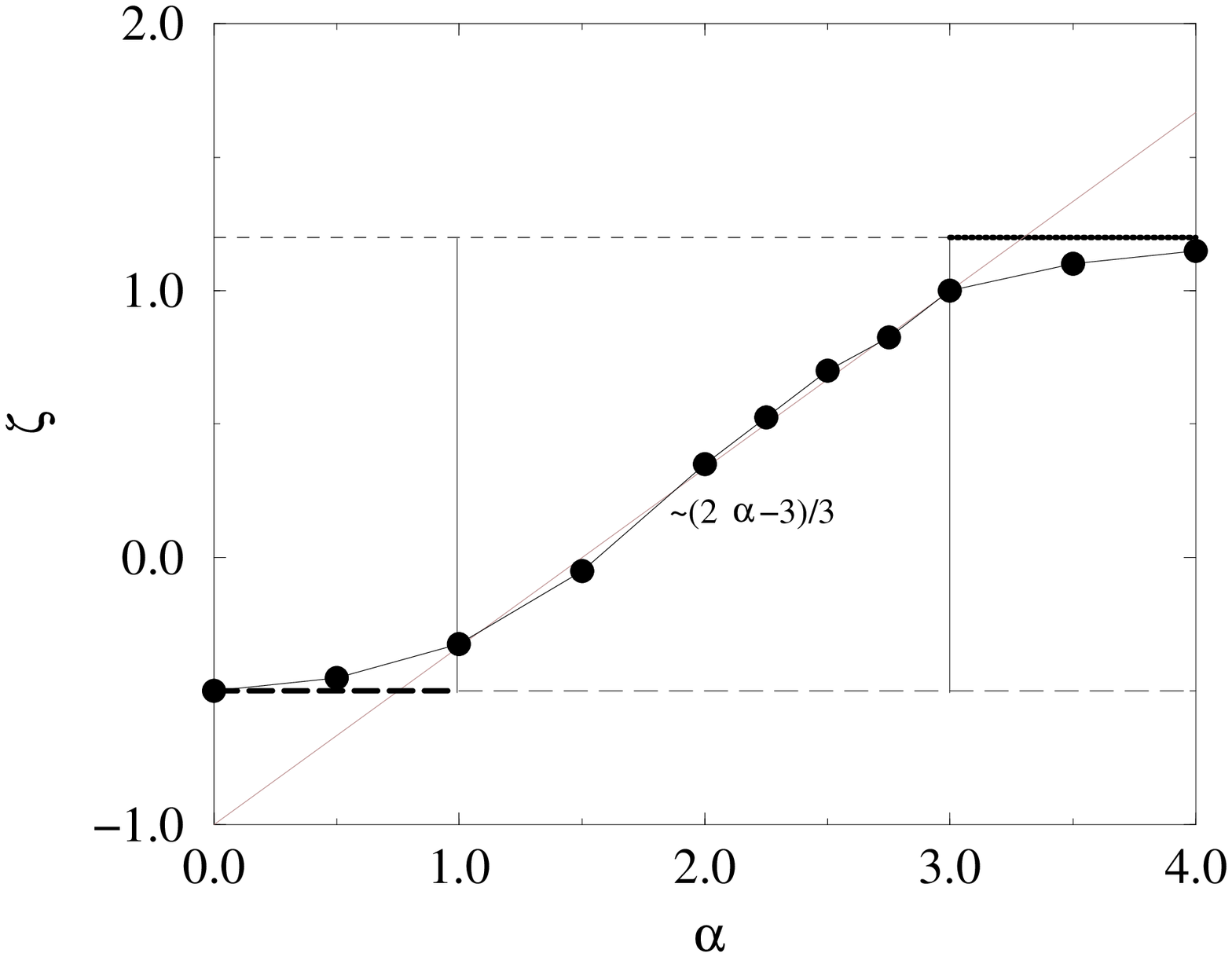}
\begin{center}Figure 5\end{center}
\end{figure} 
\begin{figure}[hbt]
\centerline{
\epsfig{file=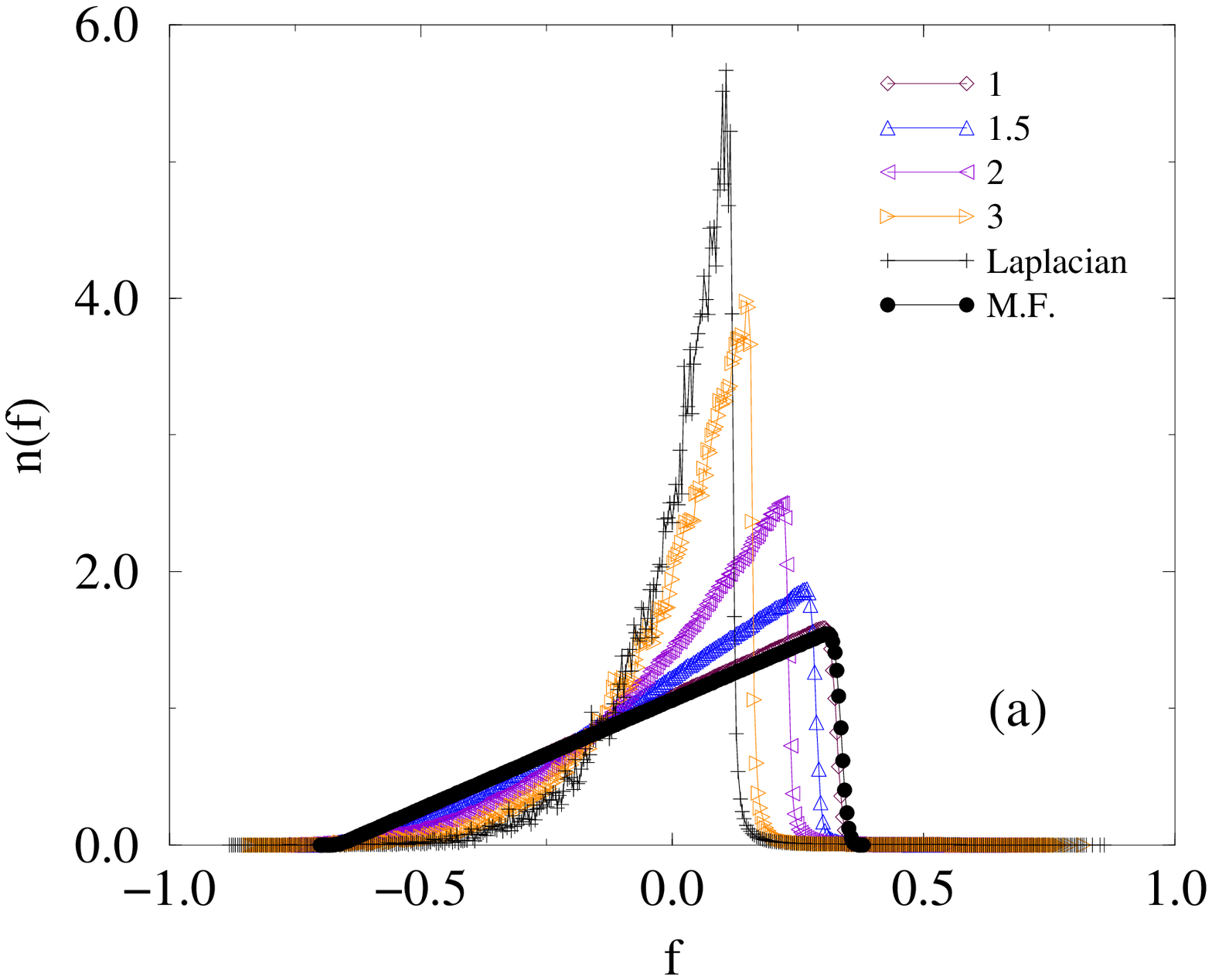,width=15cm,height=10cm}
}
\centerline{
\epsfig{file=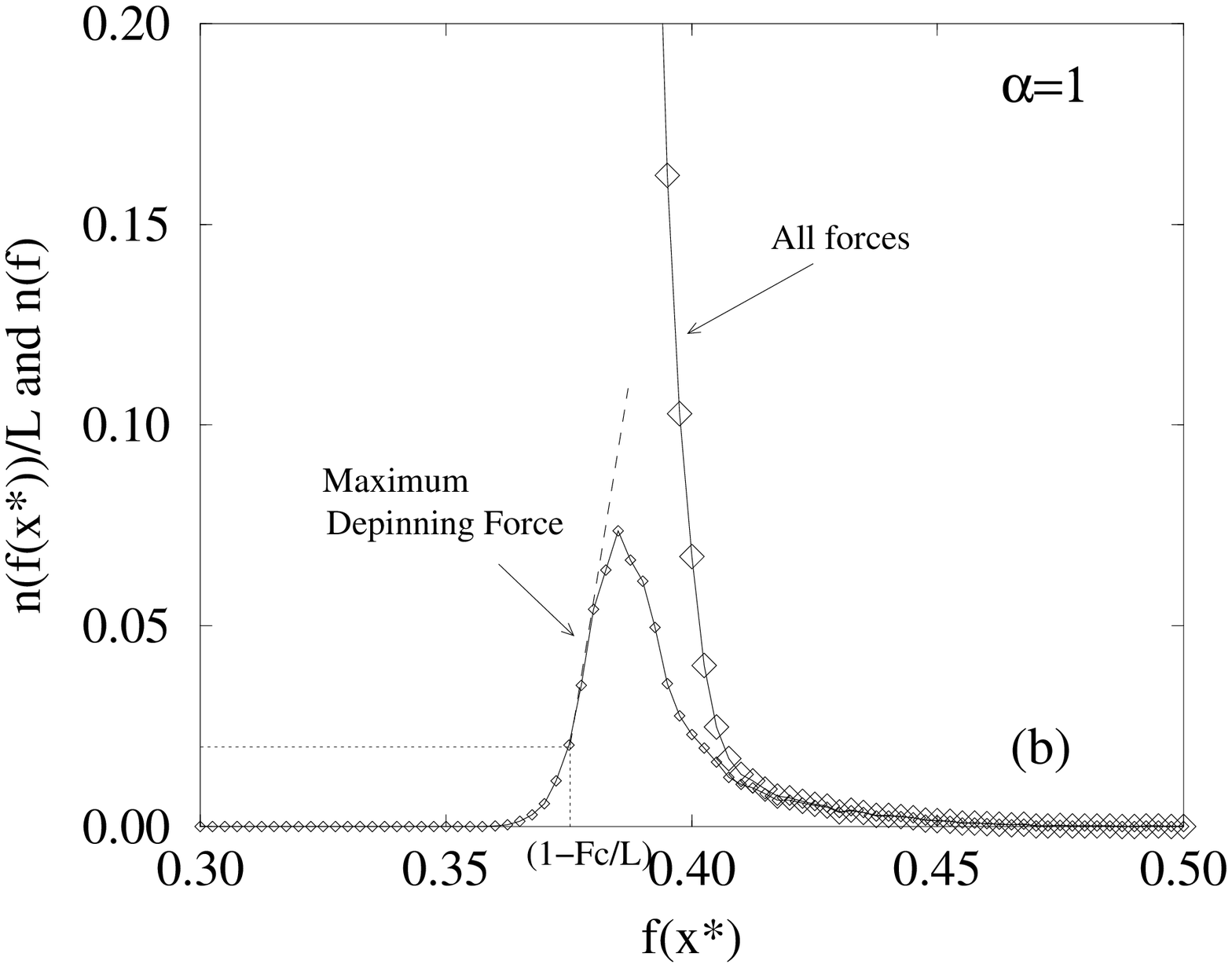,width=15cm,height=10cm}
}
\begin{center}Figure 6\end{center}
\end{figure} 
\begin{figure}[hbt]
\epsfxsize=\columnwidth\epsfbox{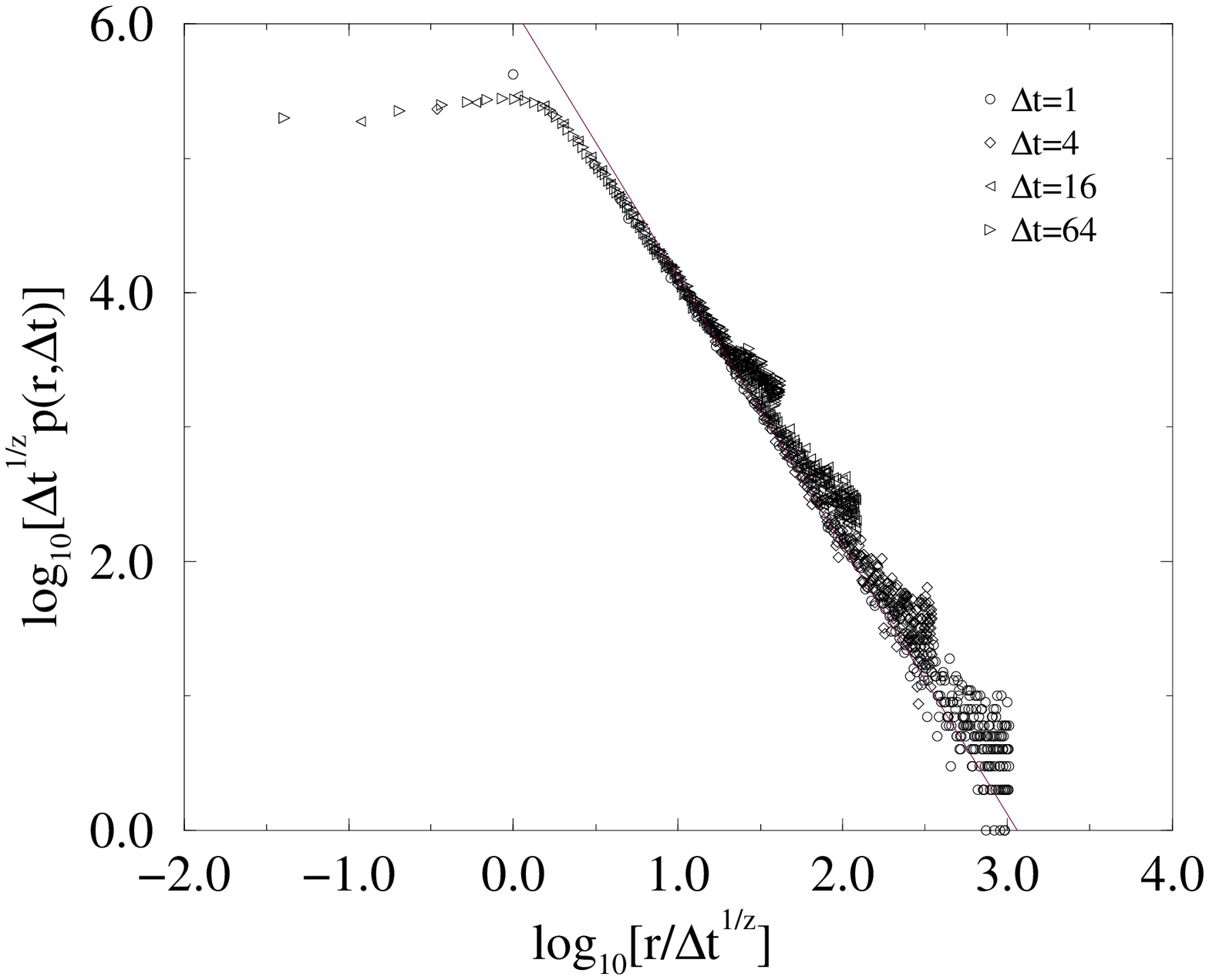}
\begin{center}Figure 7\end{center}
\end{figure} 
\begin{figure}[hbt]
\centerline{
\epsfig{file=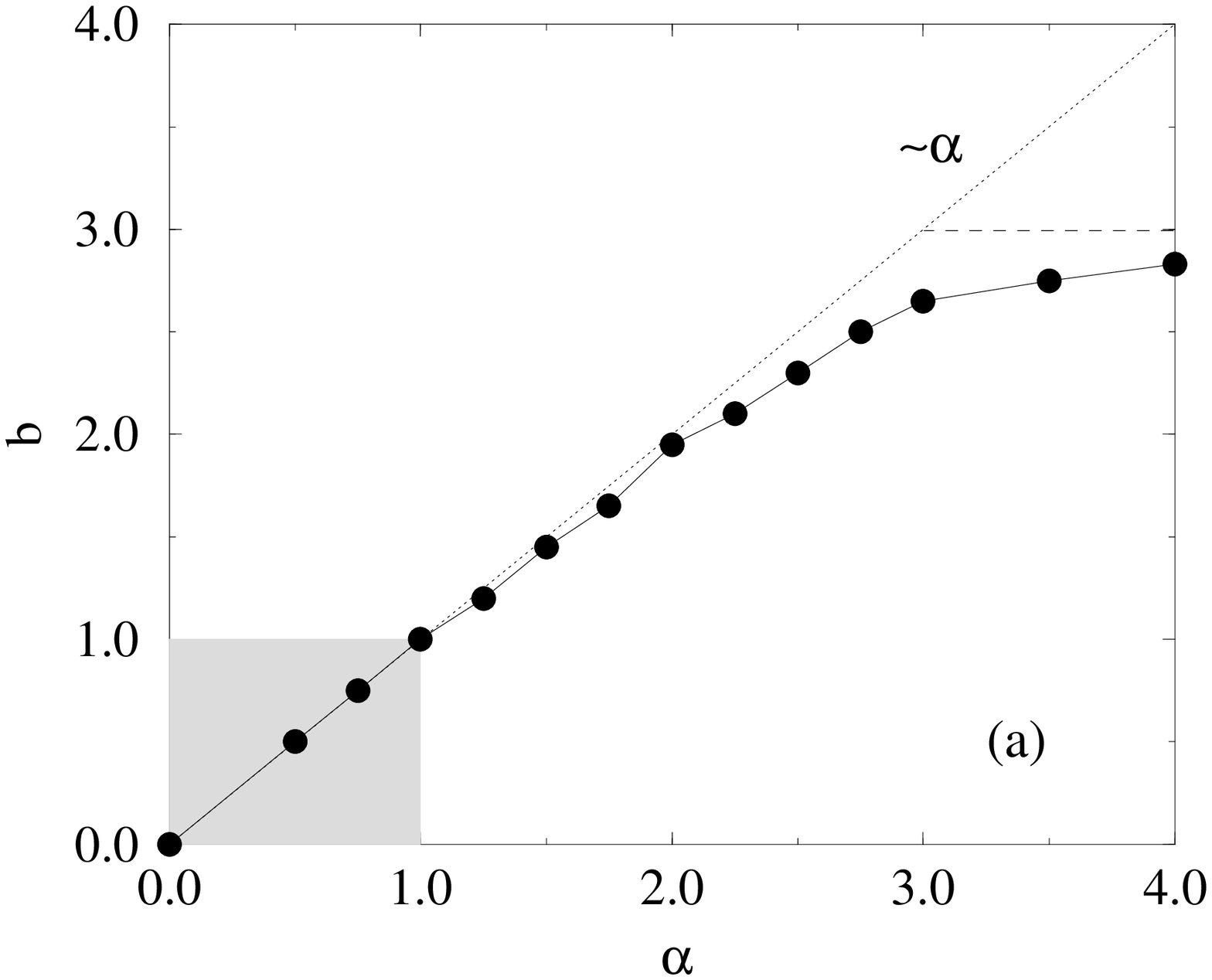,width=15cm,height=10cm}
}
\centerline{
\epsfig{file=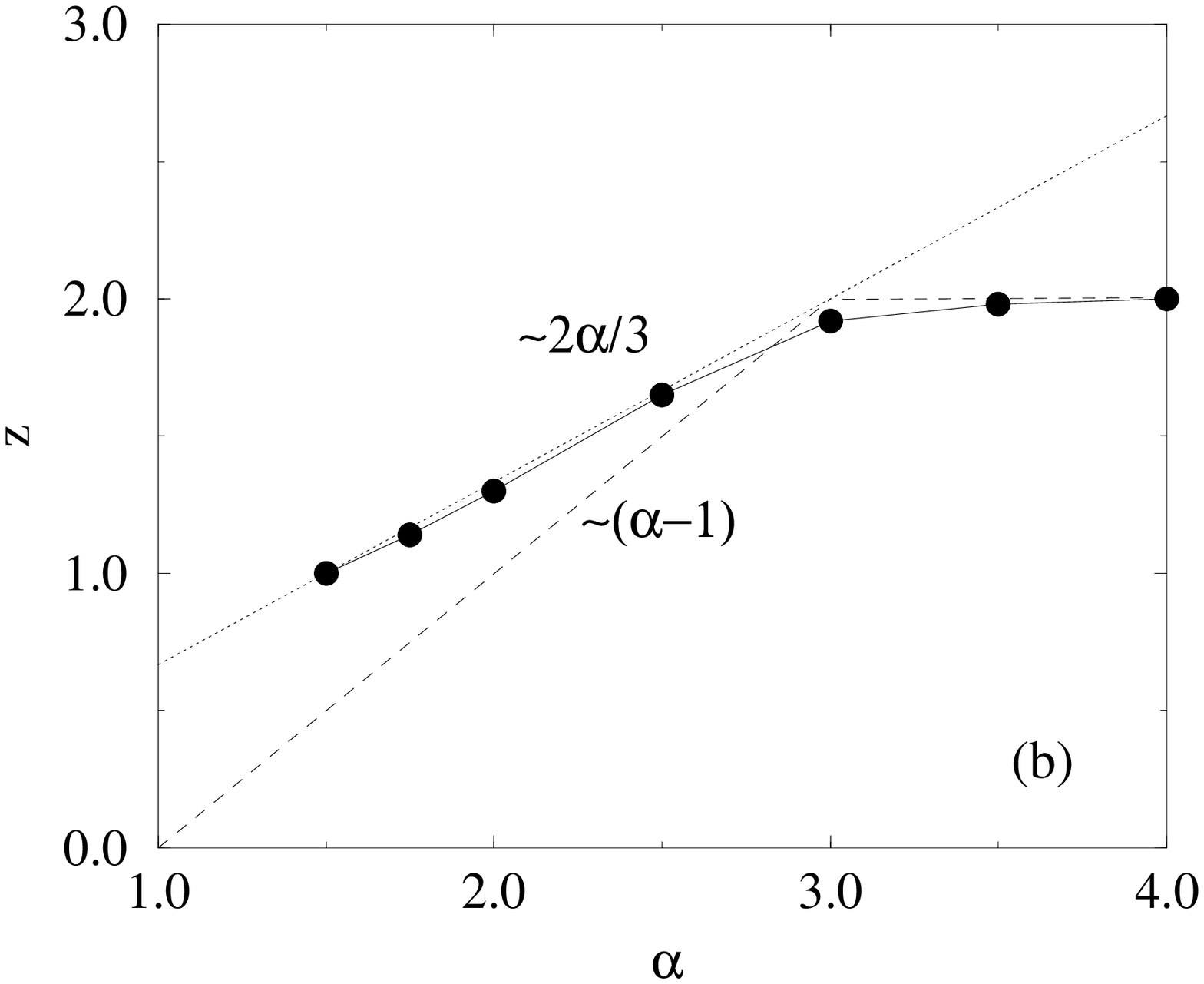,width=15cm,height=10cm}
}
\begin{center}Figure 8\end{center} 
\end{figure} 
\begin{figure}[hbt]
\epsfxsize=\columnwidth\epsfbox{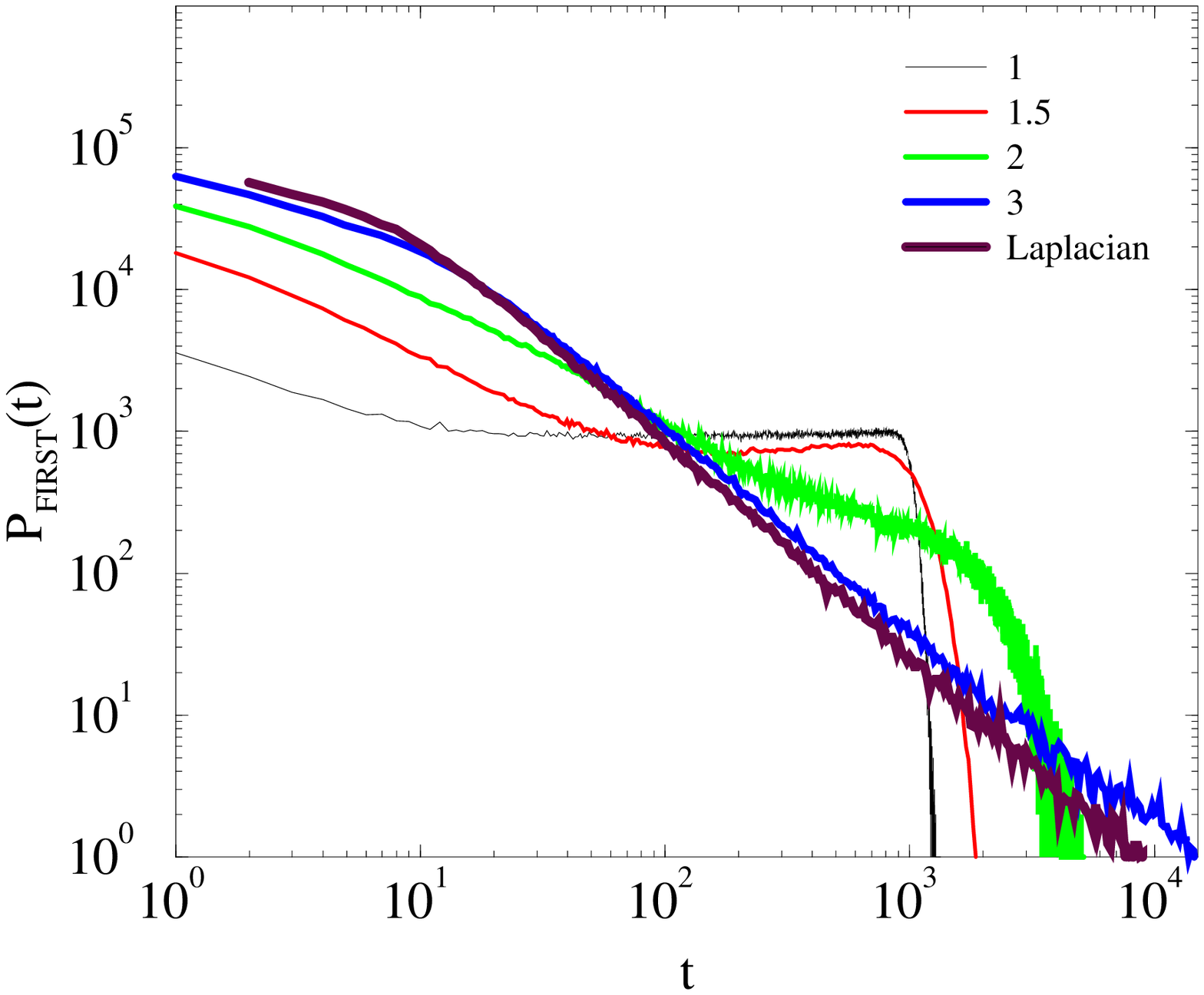}
\begin{center}Figure 9\end{center} 
\end{figure} 
\begin{figure}[hbt]
\epsfxsize=\columnwidth\epsfbox{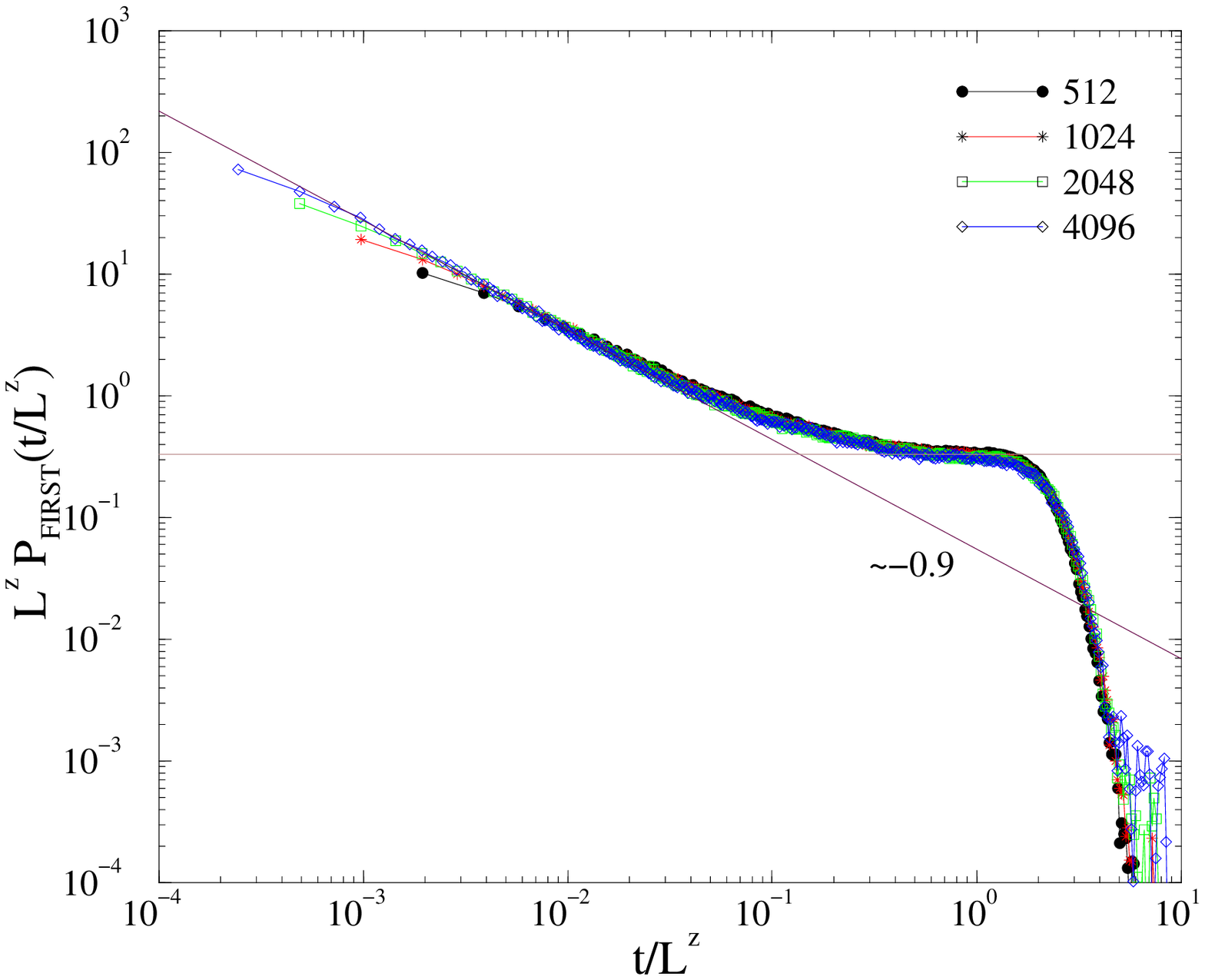}
\begin{center}Figure 10\end{center} 
\end{figure} 
\begin{figure}[hbt]
\epsfxsize=\columnwidth\epsfbox{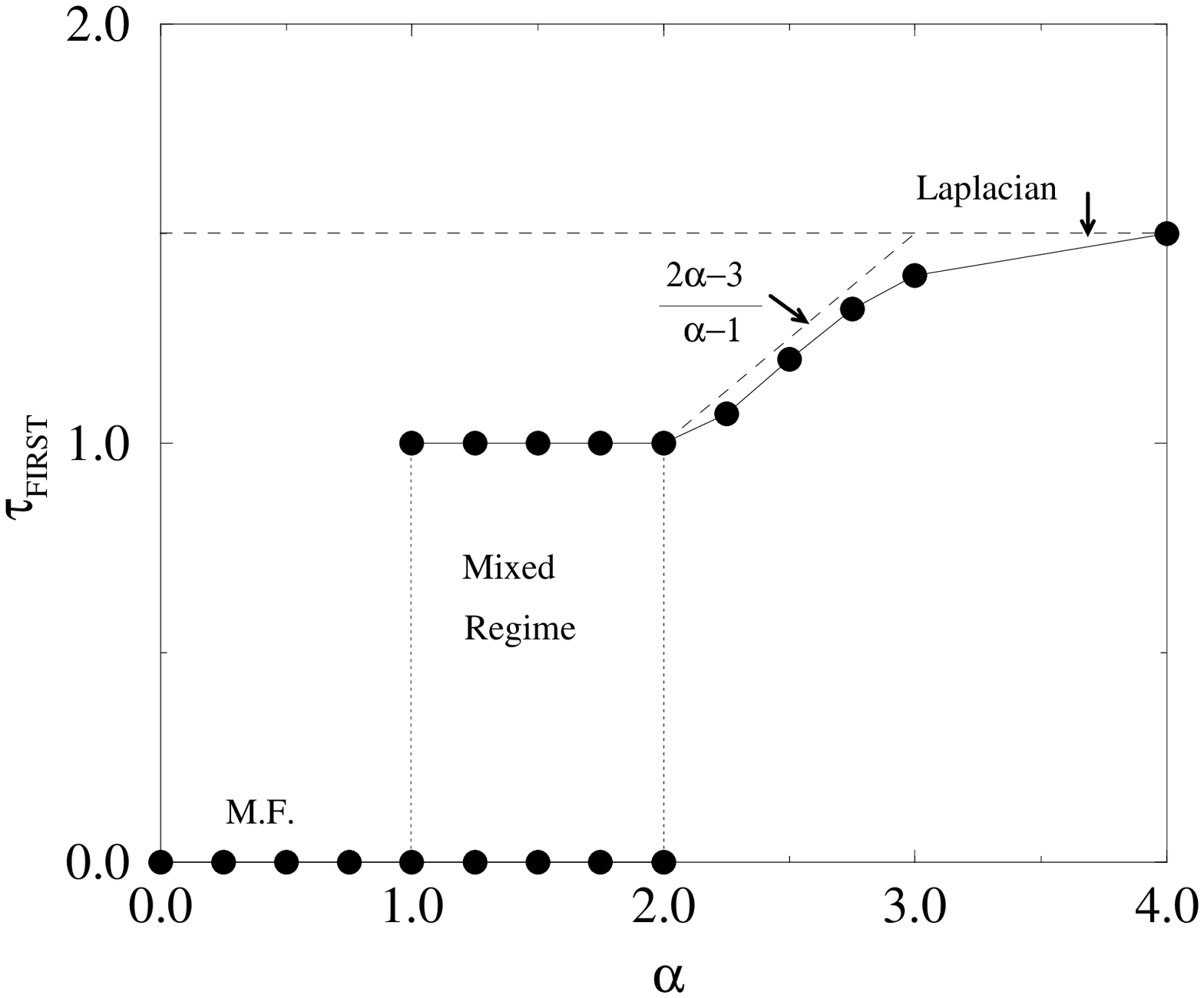}
\begin{center}Figure 11\end{center} 
\end{figure} 
\begin{figure}[hbt]
\epsfxsize=\columnwidth\epsfbox{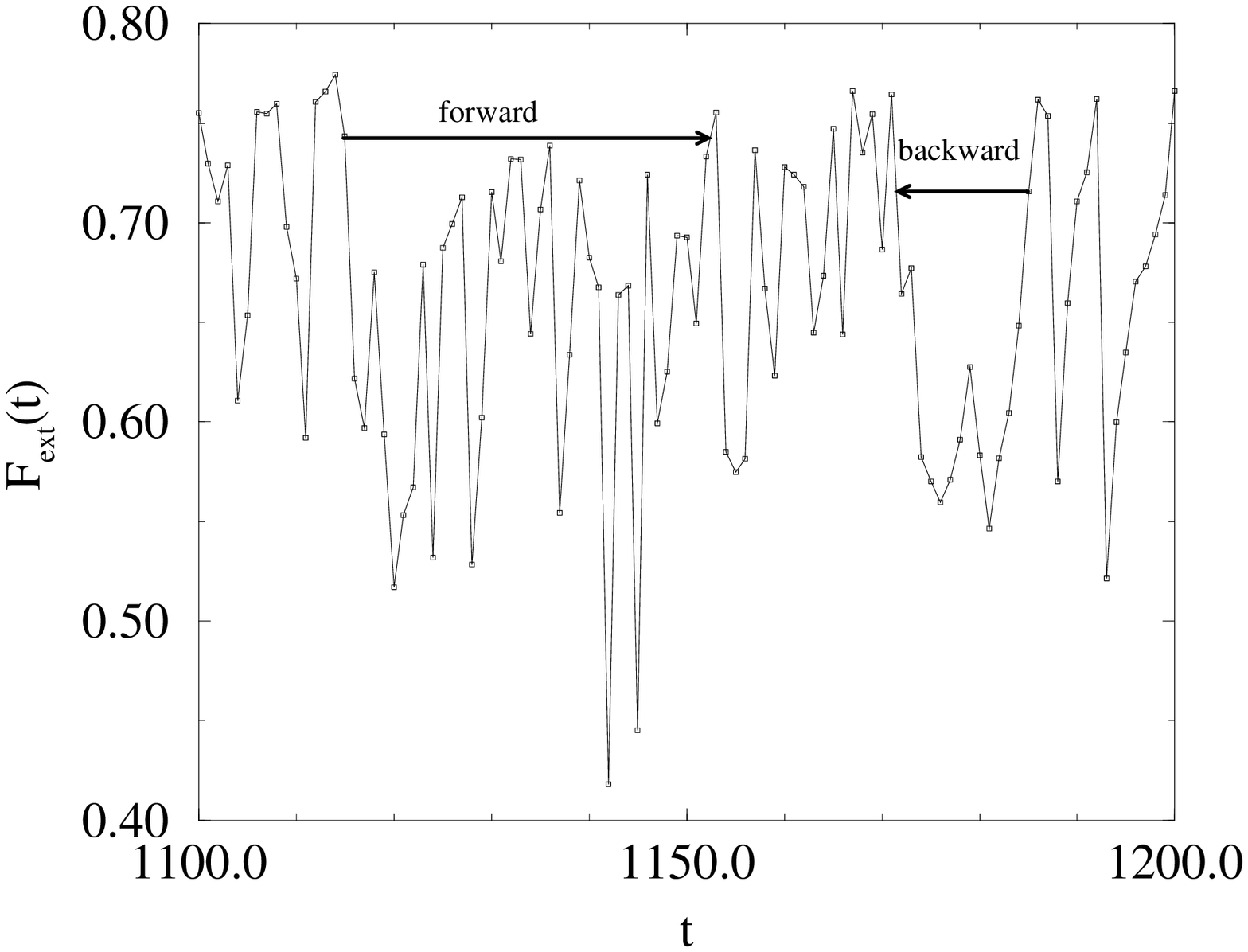}
\begin{center}Figure 12\end{center} 
\end{figure} 
\begin{figure}[hbt]
\epsfxsize=\columnwidth\epsfbox{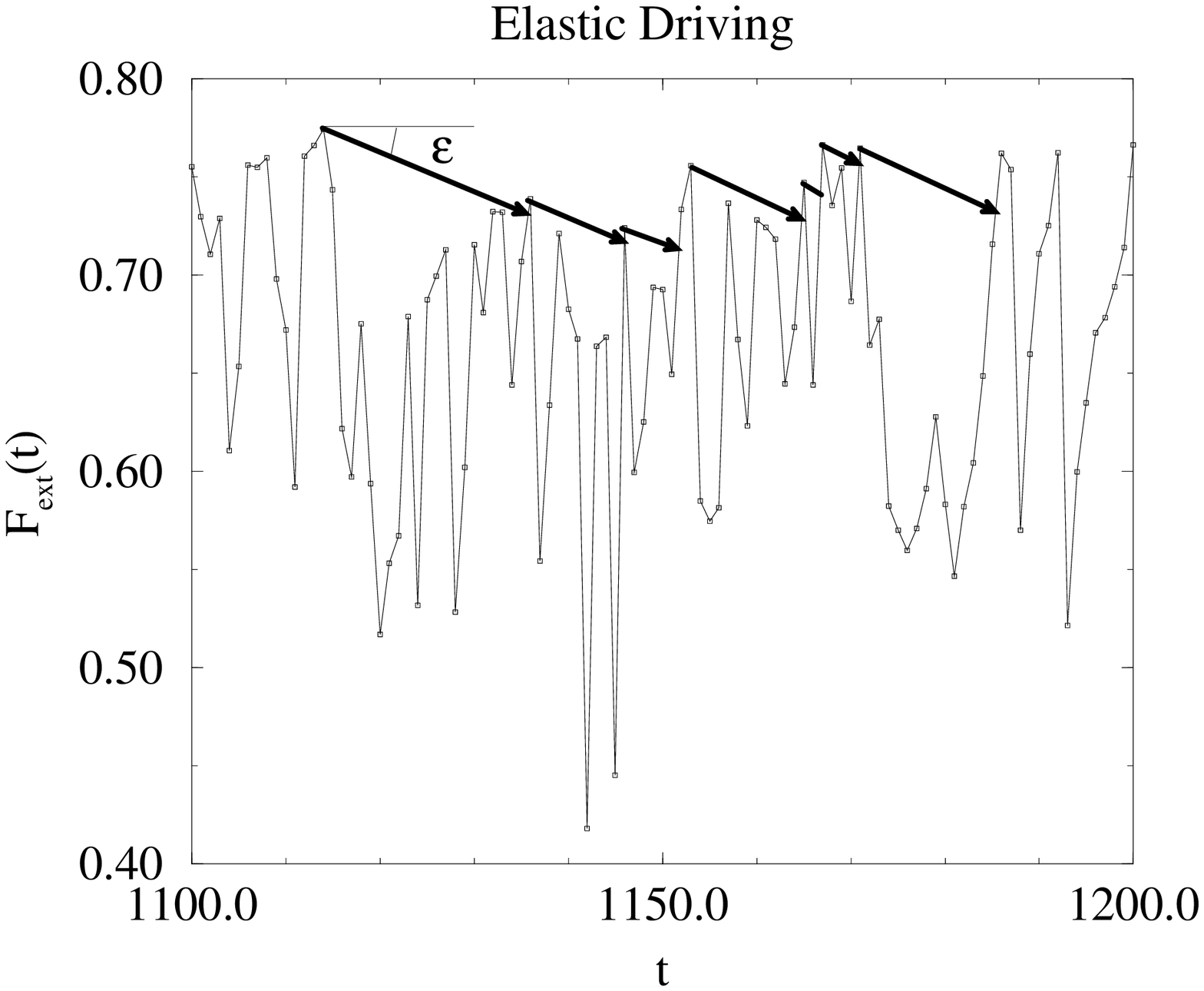}
\begin{center}Figure 13\end{center} 
\end{figure} 
\begin{figure}[hbt]
\epsfxsize=\columnwidth\epsfbox{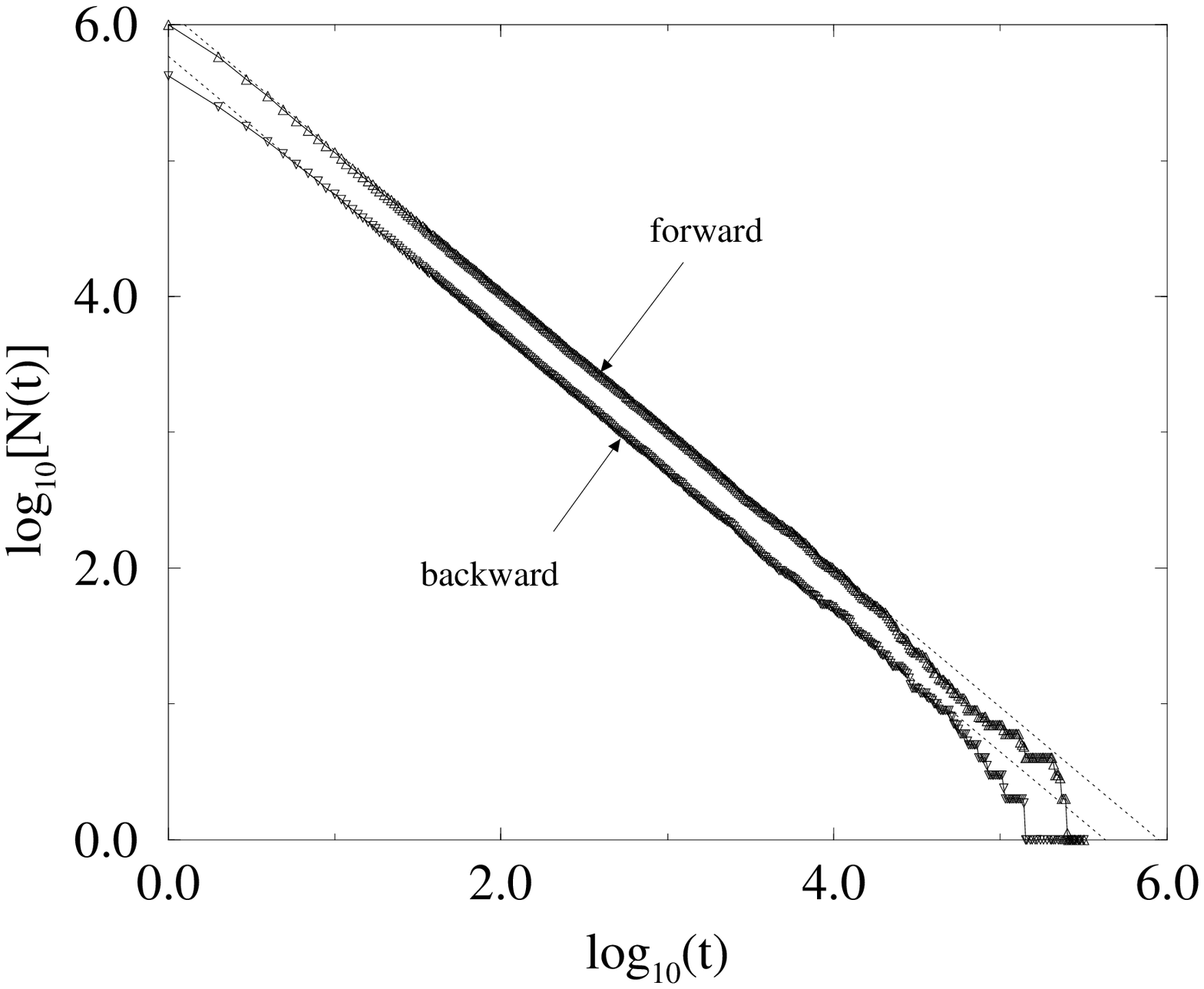}
\begin{center}Figure 14\end{center} 
\end{figure} 
\begin{figure}[hbt]
\epsfxsize=\columnwidth\epsfbox{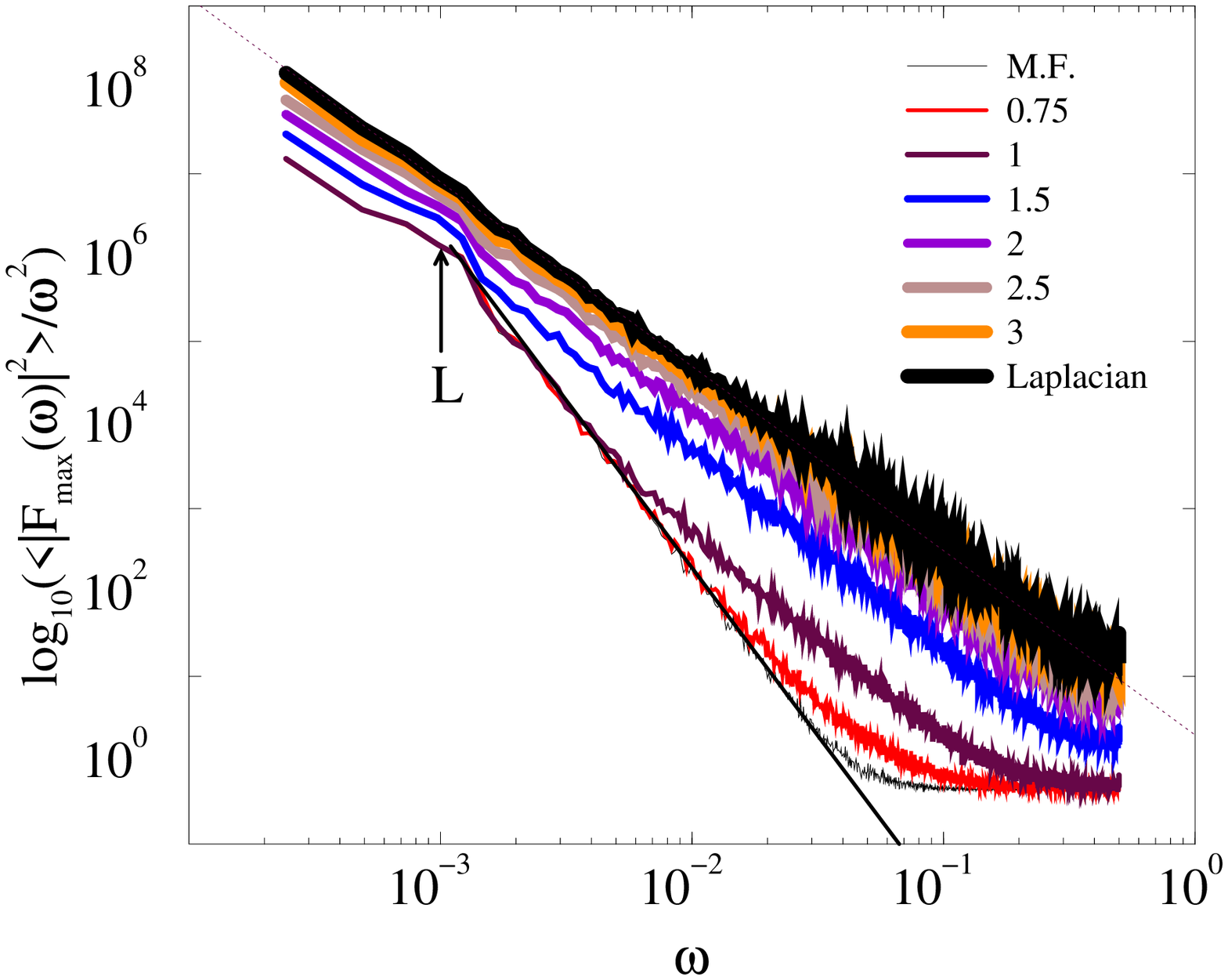}
\begin{center}Figure 15\end{center} 
\end{figure} 
\begin{figure}[hbt]
\epsfxsize=\columnwidth\epsfbox{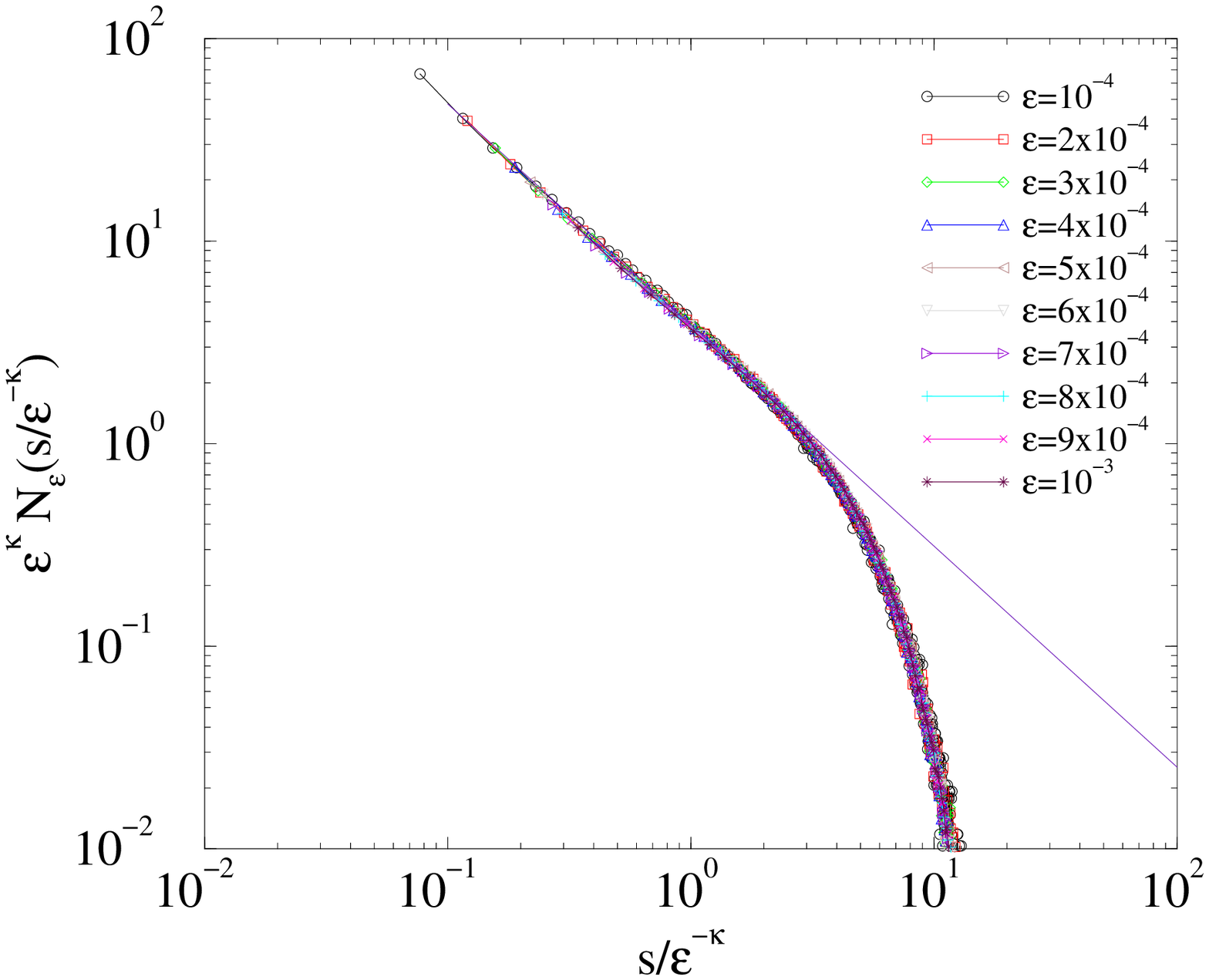}
\begin{center}Figure 16\end{center} 
\end{figure} 
\begin{figure}[hbt]
\centerline{\epsfig{file=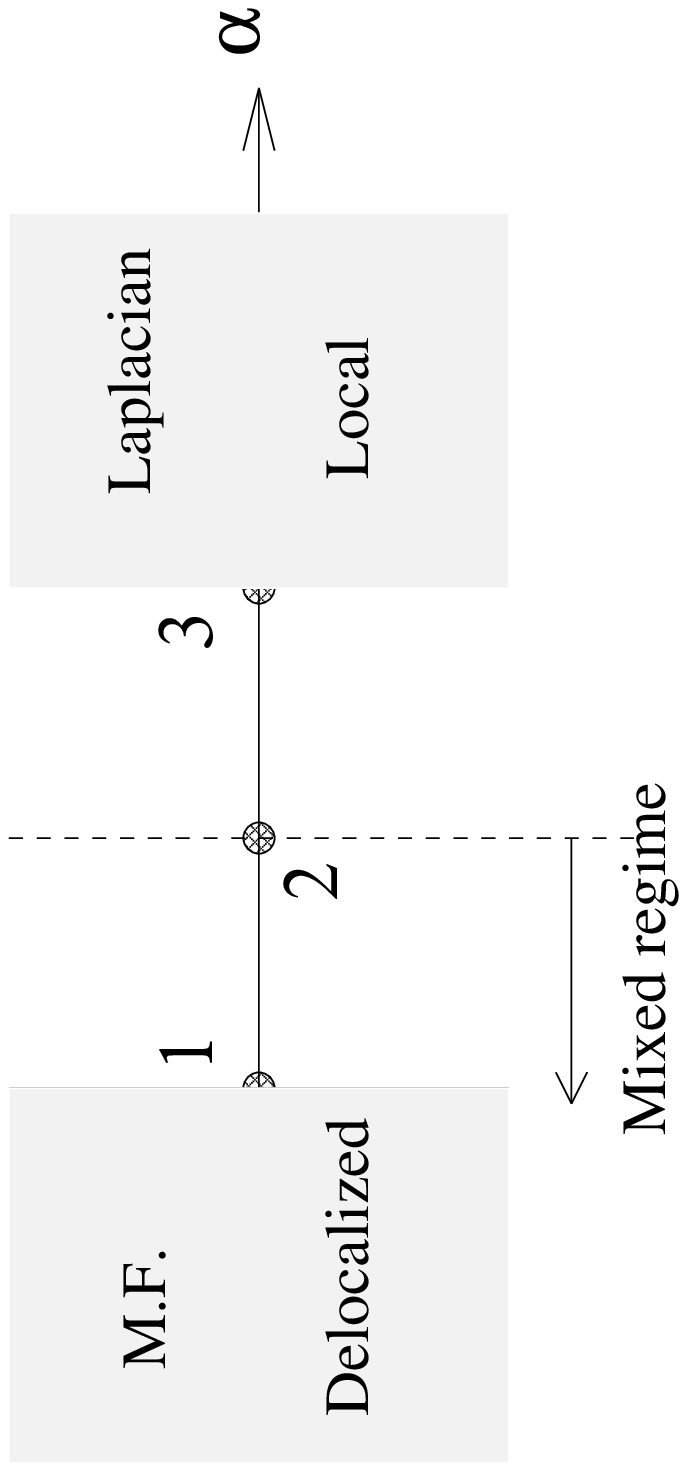,width=7cm,height=13cm,angle=-90}}
\begin{center}Figure 17\end{center} 
\end{figure} 

\begin{references}

\bibitem{Larkin} A. I. Larkin and Yu. N. Ovchinnikov, J. Low
  Temp. Phys., {\bf 34}, 409 (1979).
For a review, see T. Giamarchi and P. LeDoussal in {\it Spin Glasses and
  Random fields}
edited by A. P. Young, World Sci. ( Singapore,1997).

\bibitem{Zapp} S. Zapperi, P. Cizeau, G. Durin and H. E. Stanley, 
Cond-Mat/9803253.

\bibitem{Fisher} D. S. Fisher, Phys. Rev. B, {\bf 31}, 1396 (1985).

\bibitem{Bouchaud} J. P. Bouchaud, E. Bouchaud, G. Lapasset and J. Planes,
  Phys. Rev. Lett., {\bf 71}, 2240 (1993).

\bibitem{Caroli} C. Caroli and Ph. Nozieres in {\it The physics of
    sliding friction} edited by B. N. J. Persson, Vol.311 of NATO
  Advanced Study (Kluwer, Dordrecht,1996).

\bibitem{Mazur} P. Mazur and I. Oppenheim, Physica, {\bf 50},241 (1970).

\bibitem{Joanny} J. F. Joanny and P. G. de Gennes, J. Chem. Phys., {\bf
    81}, 552 (1984).

\bibitem{Schmitt} J. Schmittbuhl, S. Roux, J.-P. Vilotte and K.J. M\aa
  l\o y,
  Phys. Rev. Lett., {\bf 74}, 1787 (1995).

\bibitem{Gao} H.Gao and J.R.Rice, J. of Appl. Mech., {\bf 56}, 828 (1989).

\bibitem{Landau} L. Landau and E. Lifchitz, Th\'eorie de l'\'elasticit\'e, Mir Ed. (1990).

\bibitem{Lee} P. A. Lee and H. Fukuyama, Phys. Rev. B, {\bf 17}, 542 (1978).

\bibitem{Fermi} C. Castellani, C. Di Castro and A. Maccarone,
  Phys. Rev. B, {\bf
    55}, 2676 (1997).

\bibitem{Webman} O. Zik, E. Moes, Z. Olami and I. Webman, Europhys. Lett.,
  {\bf 38}, 509 (1997).

\bibitem{Paterson} A. Paterson, Annales de Physique, {\bf 21}, 337
  (1996).

\bibitem{FisherII} T. Nattermann, S. Stepanow, L. H. Tang and
  H. Leschhorn, J. Phys. II (France), {\bf 2}, 1483
  (1992); H. Leschhorn, T. Nattermann, S. Stepanow, L. H. Tang
  Cond-Mat/9603114; O. Narayan and D. S. Fisher, Phys. Rev. B, {\bf
    48}, 7030 (1993).

\bibitem{cont} T. Nattermann and L. H. Tang, Phys. Rev. A, {\bf 45},
  7156 (1992); E. Medina, T. Hwa, M. Kardar and Y. C. Zhang,
  Phys. Rev. A, {\bf 39}, 3053 (1989); K. Sneppen, Phys. Rev. Lett.,
  {\bf 69}, 3539 (1992); M. Kardar, G. Parisi, Y. C. Zhang,
  Phys. Rev. Lett., {\bf 56}, 889 (1986). For a review, see
  T. Halpin-Healy and Y. C. Zhang, Physics Reports, {\bf 254}, 215 (1995).

\bibitem{Tang} A. Tanguy and S. Roux, Phys. Rev. E, {\bf 55}, 2166, (1997).

\bibitem{Heslot} F. Heslot, T. Baumberger, B. Perrin, B. Caroli and
  C. Caroli, Phys. Rev. E, {\bf 49}, 4973 (1994).

\bibitem{WW} D. Wilkinson and J. F. Willemsen, J. Phys. A, {\bf 16},
  3365 (1983).

\bibitem{HR(chap.5-6)} H. J. Herrmann and S. Roux in {\it Statistical
    models for the fracture of disordered media}, Random Materials and
  Processes, Series Editors H. E. Stanley and E. Guyon (Elsevier
  Science Publisher, North-Holland, 1990).

\bibitem{BS} P. Bak and K. Sneppen, Phys. Rev. Lett., {\bf 71}, 4083 (1993).
 
\bibitem{Hansen} S. Roux and A. Hansen, J. Phys I France, {\bf 4}, 515
  (1994).

\bibitem{Maslov} M. Paczuski, S. Maslov and P. Bak, Phys. Rev. E, {\bf
    53}, 414 (1995).

\bibitem{Flyvbjerg} H. Flyvbjerg, K. Sneppen and P. Bak,
  Phys. Rev. Lett., {\bf 24}, 4087 (1993).

\bibitem{SchmittII} J. Schmittbuhl and K. J. M{\aa}l{\o}y,
  Phys. Rev. Lett., {\bf 78}, 3888 (1997).

\bibitem{bouchaudII} P. Daguier, B. Nghiem, E. Bouchaud and F. Creuzet,
  Phys. Rev. Lett., {\bf 78}, 1062 (1997).

\bibitem{Cieplack} M. \v Cieplack and M. O. Robbins, Phys. Rev. Lett.,
  {\bf 60}, 2042 (1988). M. \v Cieplack and M. O. Robbins,
  Phys. Rev. B, {\bf 41}, 11508 (1990). 

\bibitem{Rolley} E. Rolley, C. Guthmann, R. Gombrowicz and V. Repain, preprint (1997).

\bibitem{Joanny2} M.O. Robbins and J.F. Joanny, Europhysics Lett., {\bf 3}, 729 (1987).

\bibitem{tangroux} A. Tanguy, PhD Thesis, Universit\'e Paris VII (1998).

\bibitem{Furuberg} L. Furuberg, J. Feder, A. Aharony and T. Jossang,
  Phys. Rev. Lett.,
  {\bf 61}, 2117 (1998).
 
\bibitem{Torcini} A. Torcini and S. Lepri, Phys. Rev. E, {\bf 55},
  R3805 (1997).

\bibitem{RGuyon} S. Roux and E. Guyon, J. Phys. A, {\bf 22}, 3693 (1989).

\bibitem{Maslov_alone} S. Maslov, Phys. Rev. Lett., {\bf 74}, 562 (1995).

\bibitem{Raphael} E. Rapha\"el and P. G. de Gennes, J. Chem. Phys., {\bf 90}, 
7577 (1989).

\end{references}
\end{document}